\documentclass[aps,superscriptaddress,showpacs,nofootinbib,eqsecnum,prd,notitlepage]{revtex4} 
 
\usepackage{epsf}  
\usepackage{eucal} 
\usepackage{amssymb}
\usepackage{amsmath} 
\usepackage{graphicx} 
\usepackage{bm}
\usepackage{dcolumn} 
\usepackage{epsfig} 
\usepackage{multirow}
\usepackage{hyperref}
\usepackage{color}
\usepackage{float}
\usepackage{bm} % Bold math
 
\begin{document}

\title{Phase diagram of the magnetized planar Gross-Neveu model
beyond the large-$N$ approximation}

\author{Jean-Lo\"{\i}c Kneur} \email{jlkneur@univ-montp2.fr}
\affiliation{CNRS, Laboratoire Charles Coulomb UMR 5221, F-34095,
  Montpellier, France}
 \affiliation{Universit\'e Montpellier 2,
  Laboratoire Charles Coulomb UMR 5221,
F-34095, Montpellier, France}

\author{Marcus Benghi Pinto} \email{marcus@fsc.ufsc.br}
\affiliation{Departamento de F\'{\i}sica, Universidade Federal de Santa
  Catarina, 88040-900 Florian\'{o}polis, Santa Catarina, Brazil}

\author{Rudnei O. Ramos} \email{rudnei@uerj.br} \affiliation{Departamento de
  F\'{\i}sica Te\'orica, Universidade do Estado do Rio de Janeiro, 20550-013
  Rio de Janeiro, RJ, Brazil}

\begin{abstract}

The phase diagram and thermodynamic properties of the
(2+1)-dimensional Gross--Neveu model are studied in the presence of a
constant magnetic field. The optimized perturbation theory (OPT) is
used to obtain results going beyond the large-$N$ approximation. The
free energy and the complete phase diagram of the model, in terms of
temperature, chemical potential and magnetic field are obtained and
studied in details.  We find that some of the main qualitative changes
induced by the OPT finite $N$ corrections concern the region of
intermediate to high chemical potentials where this approximation adds
a term proportional to  $\lambda \langle \psi^+ \psi \rangle^2/N$ to
the free energy. Then, depending on the sign of $\lambda$ (relative to 
the critical coupling), and magnitude of the magnetic field, we observe a 
weakening (when $\lambda <0$) or enhancement (when $\lambda > 0$)  of the 
chiral broken region in the magnetized fermionic system. By comparing the 
results from the OPT and the large-$N$ approximation, we conclude that 
finite $N$ effects favor the phenomenon of inverse magnetic catalysis 
when the coupling constant is negative. We show that with the OPT
the value of the coexistence chemical potential at vanishing
temperature tends to decrease for large values of the magnetic field. 
This is opposite
to what is seen in the large-$N$ approximation, where for large magnetic
fields the coexistence chemical potential starts again to increase. 
Likewise, at finite temperature, the value of the chemical potential 
at the tricritical point also decreases with the magnetic field in the
OPT case. Consequently,  the shape of the
phase diagrams predicted by the OPT and by the large-$N$ approximation
look very different in the presence of high magnetic fields.
{}Finally, for small values of magnetic field and temperature,  we
identify the presence of possible intermediate nonchiral phase
transitions when varying the chemical potential. We show that these
phenomena are not an artifact of the large-$N$ approximation and that
they also occur within the OPT framework. These  intermediate
transitions are interpreted to be a consequence of the  de Haas--van
Alphen oscillations.  We also explain why this  type of phenomenon can
happen in general for negative couplings but not for positive
couplings.   

\end{abstract}


\maketitle

%%%%%%%%%%%%%%%%%%%%%%%%%%%%%%%%%%%%%%%%%%%%%%%%%%%%%%%%%%%%%%%%%%%%%%%

\section{Introduction}

{}Four-fermion theories~\cite{rose} find applications in several areas
of physics, from condensed  matter systems (for example in models for
polymers, high temperature superconductors, etc) to high energy
physics, most notably as effective models for QCD, 
like the Nambu--Jona-Lasinio (NJL)  model~\cite{njl} and its
variants, including the Gross--Neveu (GN) model~\cite{gn}.  These
theories are typically employed in  the study of  chiral transition,
either for the discrete form of the symmetry, $\psi \to \gamma_5
\psi$, or in the continuous form, $\psi \to \exp(i \alpha \gamma_5)
\psi$. The general interest is to understand how the chiral transition
pattern is affected by  external control parameters such as the
temperature ($T$) and the chemical potential ($\mu$). It is also well
known that, apart from these control parameters, the presence of an
external magnetic field may  impact significantly on the phase
transition patterns.   

So far   we have a reasonable understanding of how  chiral symmetry is
affected by $T$, $\mu$ as well as  by the presence of a magnetic field
$B$.  Mostly, this understanding is acquired  at the mean-field
level~\cite{varios} for both the GN and the NJL models. Recently, the
understanding of how an external $B$ field affects the symmetry
aspects  of these four-fermion theories turned out to be a question of
general interest for the following reasons. {}First, the interaction
of fermions with an external $B$ field is expected to be associated
with phenomena such as the metal-to-insulating phase transition in
semiconductors~\cite{mit}, quantum Hall effects~\cite{hall} and the
transport properties in superconductors~\cite{supercond},  just to
mention a few phenomena in the context of condensed matter physics,
while in the high energy physics domain the effects of a magnetic
background is important to the physics of compact stellar objects
\cite{magnetars}, heavy-ion collisions at 
the Relativistic Heavy Ion Collider and at the LHC \cite{kharzeev09} 
and in the physics of the early universe \cite{universe}, which are 
situations where high intensity fields, $B\sim 10^{17}-10^{20} \,G$, 
are expected to be present or produced (for a recent review, see e.g. 
Ref.~\cite{Fukushima:2011jc} and references therein).

Of topical interest is to understand how a magnetic field will affect
phase transitions, since they  can induce dynamical symmetry breaking,
or magnetic catalysis~\cite{Klimenko,Gusynin}. Within fermionic
systems\footnote{Though magnetic catalysis is usually related to the
  physics of fermionic systems, it can also work for bosonic scalar
  fields as well~\cite{magboson},  by enhancing the ordered phase in
  the presence of an external magnetic field.}, magnetic catalysis
refers to the generation of a mass gap for the fermions at any finite
interaction strength, leading explicitly to chiral symmetry
breaking \footnote {Chiral symmetry breaking can also arise in free
  fermionic systems as a consequence of the {\it quantum anomalies}
  although, in this case, it does not produce a mass term for the
  fermions, whose spectrum remains unchanged~\cite{hall}.}. 

In this work, we apply the optimized perturbation theory (OPT)
\cite{optmethod} (see Ref. \cite {early} for early works on this
subject)  to the GN model in  $2+1$ dimensions  and investigate how
dynamical chiral symmetry breaking (CSB) is affected by the presence
of a magnetic field,  comparing our results with those obtained in the
large-$N$ (LN) approximation, which is equivalent to the well known
mean field approximation (MFA).  The OPT has already established
itself as a powerful method in dealing with  critical theories. {}For
example, in the Bose--Einstein condensation case this method and its
different variations have provided some of the most precise analytical
results for the shift in the critical temperature for weakly
interacting homogeneous Bose gases~\cite{bec}. Other applications to
condensed matter situations include a precise evaluation of the
critical density for polyacetylene \cite{poly}.  Also, when extended
by hard-thermal loops, the method was successful in predicting QCD
thermodynamical properties at the three-loop
level~\cite{mike}. Improved by the renormalization group, and
inspired by similar properties~\cite{rgopt1} in the Gross--Neveu model,
a variation of the OPT has been recently used in the evaluation of
$\Lambda_{\rm MS}^{\rm QCD}$~\cite{rgoptqcd1} and
$\alpha_S$~\cite{rgoptqcd2}, where the stability and  convergence at
higher orders of this renormalization group OPT form was demonstrated. 
{}For the present
application it is worth mentioning that the OPT was instrumental in
the determination of the phase diagram of the massless GN model in 2+1
dimensions at finite $T$ and $\mu$ in the absence of magnetic
fields~\cite{prdgn3d}. In this case, the LN approximation predicts
that the whole $T-\mu$ plane is dominated by a second-order phase
transition, except at $T=0$, where a first-order phase transition is
predicted to occur. But,  Monte--Carlo numerical
simulations~\cite{kogut,hands} have indicated that a first-order
transition line should appear at the low-$T$ and high-$\mu$ region,
terminating in a tricritical point at intermediate values of $T$ and
$\mu$.  However, no precise location for this tricritical point was
possible to be given. This situation has been changed when the
complete phase diagram for the model was studied in the context of the
OPT method and the precise location of the tricritical point
determined for any value of $N$~\cite{plbgn3d}. The two-flavor NJL
model in $3+1$ dimensions with physical quark masses has also been
treated with the OPT at finite $T$ and $\mu$ in the absence of
magnetic fields~\cite{njl2F}. The main outcome was that the $1/N$
corrections brought in by this approximation  generate the effects of
a repulsive vector channel (absent in the original  Lagrangian), which
weakens the first-order transition line and locates the critical end
point at temperature values that are smaller than the ones predicted
by the LN approximation.  A detailed discussion about the physical
nature of the OPT $1/N$  corrections in the simplified Abelian NJL
model context was  recently carried out in Ref.~\cite{ijmpe}.

In the massless GN model, the LN approximation  predicts that the
critical temperature, which signals that chiral symmetry has been
restored through a second-order phase transition, increases with $B$
at vanishing fermionic densities.  However, although the functional
renormalization group technique~\cite{Scherer:2012nn} has been
recently applied to analyze magnetic catalysis at zero temperatures
and densities,  we are not aware of evaluations that go beyond
mean field at finite $T$, $\mu$ and $B$ in the context of the GN model
in $2+1$ dimensions.  In the present work we show that at $\mu=0$ the
results obtained with the LN method and the OPT agree from the
qualitative point of view. Namely, magnetic catalysis still takes
place and the critical temperature rises with $B$.    However, at the
other extreme of the phase diagram, when $T=0$ and $\mu \ne 0$,  we
find that the OPT and the LN predict different qualitative and
quantitative behaviors as far as the  coexistence chemical potential
$\mu_c$ is concerned.  The differences are more pronounced for
negative couplings ($\lambda<0$),  where the OPT reproduces the
phenomenon of inverse magnetic catalysis (IMC)~\cite{andreas},  which
predicts the decrease of $\mu_c$ with increasing $B$, even at large
magnitude of magnetic fields. Other qualitative differences happen in
the region spanned  by intermediate to high chemical potentials, where
the OPT adds terms of the form  $\lambda \langle \psi^+ \psi
\rangle^2/N$ to the free energy, while only scalar  condensates
($\langle {\bar \psi}\psi \rangle$) are considered within the LN
approximation.  Then, depending on the sign of the four-fermion
coupling $\lambda$ and for sufficiently large values of the magnetic
field,  we observe the weakening ($\lambda < 0$) or the enhancement
($\lambda > 0$)  of   the entire CSB region in the magnetized
fermionic system.    The OPT also predicts that the value of $\mu_c$
(at $T=0$) can be smaller than the tricritical point value ($\mu_{\rm
  tric}$),  producing an important change in the shape of the phase
diagram as compared to the one generated by taking $N \to
\infty$. {}Finally, we also discuss the possibility that the  order
parameter value suffers more than one discontinuity as $\mu$ increases
when  $T=0$ and $B$ is small. Being observed by both  approximations
when $\lambda <0$,  this feature is not an artifact of the LN
approximation and can be easily explained by a  close examination of
the filling of the Landau levels.

This paper is organized as follows. In the next section we briefly
review the planar GN four-fermion model within the OPT formalism.  In
Sec.~\ref{sec3} we obtain the effective potential (or free energy
density), at first nontrivial order,  which is adequate to treat a
hot, dense and magnetized  planar four-fermion system.  Next, in
Sec.~\ref{sec4} we discuss how finite $N$ effects affect
thermodynamical quantities  such as the order parameter, the critical
quantities  and the overall shape of the phase diagram  when a
magnetic  field is present. The results  obtained from the OPT are
contrasted with those produced by the LN approximation in all the
cases we have analyzed.  {}Finally, in Sec.~\ref{sec5}, we give our
concluding remarks.  Two appendices are also included to show and
clarify some technical aspects.

%%%%%%%%%%%%%%%%%%%%%%%%%%%%%%%%%%%%%%%%%%%%%%%%%%%%%%%%%%%%%%%%%%%%%%%

\section{GN Model in an external constant magnetic field 
in the OPT formalism}
\label{sec2}

In the presence of an external electromagnetic potential $A_\mu$, the
GN model with fermions  with $N$ flavors, $\psi_k$ ($k=1,\ldots,N$),
is described by the Lagrangian density~\cite{gn}

\begin{equation}
{\cal L} = \bar{\psi}_{k} \left( i \not\!\partial - e \not \!\! A
\right ) \psi_{k} - m_f {\bar \psi_k} \psi_k + \frac {g^2}{2} ({\bar
  \psi_k} \psi_k)^2\;.
\label{GN}
\end{equation}
Note that a  summation over fermionic species is implicit in the above
equation with, e.g.,  $\bar{\psi}_k \psi_k = \sum_{k=1}^N \bar{\psi}_k
\psi_k$.  When $m_f=0$, which is the case considered by us here,  the
theory is invariant under the discrete chiral symmetry (CS)
transformation

\begin{equation}
\psi \to \gamma_5 \psi \,\,\,,
\end{equation}
with the gamma matrices being $4\times 4$ matrices and we follow the
representation given, e.g., in Ref.~\cite{rose} for fermions in 2+1
dimensions\footnote{Note that chiral symmetry breaking in 2+1 dimensions
requires fermion fields with four-component spinors so that the
gamma matrices are represented by  $4\times 4$ Dirac matrices 
(see e.g. Ref.~\cite{chiralsym} for details), as we implicitly consider 
in this work.}.  
A constant magnetic field $B$ along the $z$ direction,
perpendicular to the plane of the  system defined by Eq. (\ref{GN}),
can be considered by choosing a gauge where the external
electromagnetic potential is given, for example, by $A_\mu = (0,0,B
x,0)$.

The LN limit (or MFA) of the model Eq.~(\ref{GN}) is defined by
considering the four-fermion interaction as $g^2= \lambda/N$ and
taking $N\to \infty$, while keeping  $\lambda$ fixed (see, e.g.,
Ref.~\cite{coleman}).   In the following we will study the model of
Eq.~ (\ref{GN}) beyond the simplest MFA/LN approximation by employing
the OPT method.

Within the OPT framework one makes use of  a linear interpolation on
the original model in terms of a fictitious parameter, $\delta$ (used
only for bookkeeping purposes), which allows for further
expansions~\cite{optmethod}. Then, following
e.g. Refs.~\cite{prdgn2d,prdgn3d}, the interpolated GN four-fermion
theory can be expressed as

\begin{equation}
{\cal L}_{\delta}(\psi, {\bar \psi}) = \bar{\psi}_{k} \left( i
\not\!\partial - e \not \!\! A \right) \psi_{k} - (1-\delta)\, \eta \,
               {\bar \psi_k} \psi_k + \delta \frac {\lambda}{2N}
               ({\bar \psi_k} \psi_k)^2\;.
\label{GNlde}
\end{equation}
Note that at $\delta=0$ we have a theory of free fermions, while at
$\delta=1$ we recover the original theory.  We can now rewrite the
four-fermion interaction in Eq.~(\ref{GNlde}) by introducing an
auxiliary scalar field $\sigma$ in the usual way~\cite{coleman}, such
that Eq.~(\ref{GNlde}) becomes:

\begin{equation}
{\cal L}_{\delta} = \bar{\psi}_{k} \left( i \not\!\partial - e \not
\!\!A \right) \psi_{k} - \delta \sigma {\bar \psi_k} \psi_k -
(1-\delta) \, \eta \, {\bar \psi_k} \psi_k - \frac {\delta N }{2
  \lambda } \sigma^2   \;,
\label{GNdelta}
\end{equation}
where $\sigma$ and the chiral operator are related, from the
saddle-point solution for $\sigma$, by $\sigma= -(\lambda/N)
\bar{\psi}_k\psi_k$.  Renormalization issues do not arise at the level
of the approximation considered in the present application, but the
interested reader can find a comprehensive discussion in the context
of the OPT method in Ref.~\cite{prdgn3d}, for example. 

Any quantity computed from the above interpolated Lagrangian density
(\ref{GNdelta}), at some finite order in $\delta$, is dependent on the
arbitrary mass parameter $\eta$,  which also serves as an infrared
regulator. Then, after the formal mathematical manipulations
associated with the  evaluation of the relevant Green functions, one
must fix the arbitrary $\eta$ in a judicious way.  Here, as in most of
the previous works on the OPT method (see, e.g.,
Refs.~\cite{ldephi4,prdgn2d,prdgn3d}),  $\eta$ is fixed by using the
principle of minimal sensitivity (PMS). Within the PMS procedure one
requires that a physical quantity  $\Phi^{(k)}$, that is calculated
perturbatively to some $k$-th order in $\delta$, be evaluated at the
point where it is less sensitive to this mass parameter.  This
criterion then translates into the variational relation~\cite{pms}

\begin{equation} 
\frac {d \Phi^{(k)}}{d \eta}\Big |_{\bar \eta, \delta=1} = 0 \;. 
\label{PMS} 
\end{equation} 
The optimum value $\bar \eta$ that satisfies Eq.~(\ref{PMS}) must be a
function of the original parameters, including the couplings, thus
generating in that sense nonperturbative dependences in the coupling
and other parameters of the model.

%%%%%%%%%%%%%%%%%%%%%%%%%%%%%%%%%%%%%%%%%%%%%%%%%%%%%%%%%%%%%%%%%%%%%%%

\section{Effective potential for the interpolated theory}
\label{sec3}

{}Following, e.g., Ref.~\cite{prdgn3d},  the effective potential (or
free-energy density) for a constant background scalar field,
$\sigma_c$, at first order in the OPT approximation is given by

\begin{equation}
\frac{1}{N}V_{{\rm eff},\delta^1}(\sigma_c,\eta) = \delta \frac
     {\sigma_c^2}{2 \lambda} + i  \int_p  {\rm tr}\ln \left(\not \! P
     - \eta \right)+ \delta  i \int_p   {\rm tr} \frac
     {\eta-\sigma_c}{\not \! P - \eta + i \epsilon} +
     \frac{1}{N}\Delta V_{{\rm eff},\delta^1} \;,
\label{general}
\end{equation}
where $\Delta V_{{\rm eff},\delta^1}/N$ brings the first $1/N$
corrections to the effective potential. This contribution is
explicitly  given by

\begin{equation}
\frac{1}{N}\Delta V_{{\rm eff},\delta^1} = - \frac {i}{2N}  \int_p
     {\rm tr} \left [\frac {\Sigma_{\delta^1}(\eta)}{\not \! P - \eta
         + i \epsilon} \right ]\;,
\label{VN1}
\end{equation}
where $\Sigma_{\delta^1}(\eta)$  is the ${\cal O}(\delta)$
contribution to the fermion self-energy,

\begin{equation}
\Sigma_{\delta^1} (\eta) = -\delta  \frac {\lambda}{N}  i   \int_q
\frac {1}{\not \! Q - \eta+i \epsilon}\;.
\label{Sigma1}
\end{equation}
The traces in Eqs.~(\ref{general}) and (\ref{VN1}) are taken over
Dirac's  matrices only (a factor of $-1$, corresponding to a closed
fermionic loop, has already been taken into  account~\cite{root} in
the above expressions).  After taking the traces over the Dirac's
matrices and rearranging the terms, Eq.~(\ref{general}) can be written
as 

\begin{eqnarray}
\frac{1}{N}V_{{\rm eff},\delta^1}(\sigma_c,\eta) &= &\delta \frac
     {\sigma_c^2}{2 \lambda} + 2  i  \int_p \ln \left( P^2 - \eta^2
     \right)+\delta\: 4 i  \int_p \frac {\eta(\eta-\sigma_c)}{P^2 -
       \eta^2 + i \epsilon} \nonumber \\ &+& \delta  \frac{2
       \lambda}{N} \eta^2\left [i \int_p \frac {1}{P^2 - \eta^2 + i
         \epsilon} \right ]^2 + \delta \frac{2 \lambda}{N} \left [i
       \int_p \frac {P_0}{P^2 - \eta^2 + i \epsilon} \right ]^2 \;.
\label{vlde1ddim}
\end{eqnarray}
In the above expressions we are  using the notation for the momentum
integrations in $(2+1)$ dimensions, at finite temperature and chemical
potential and in the presence of a  constant magnetic field. Expressed
in terms of  sums over discrete Matsubara's frequencies and Landau
levels (LLs)  the integral measure is given by 

\begin{equation}
\int_p= \int \frac{d^3p}{(2\pi)^3} \equiv  i T \sum_{\nu =
  -\infty}^{\infty}  \frac{eB}{2\pi}  \sum_{j=0}^{\infty}
\frac{2-\delta_{j,0}}{2} \;,
\label{intp}
\end{equation}
where $\omega_\nu= (2 \nu+1)\pi T$, with $\nu=0,\pm 1,\pm 2,\ldots$,
are the Matsubara frequencies for fermions, with $T$ the
temperature. The sum over $j$ are over the Landau levels (LL), with a
density of states $eB/(2 \pi)$. The time and space components of the
momentum are $p_0 \to i(\omega_\nu - i \mu)$, where $\mu$ is the
chemical potential, while ${\bf p}^2 \to 2 j eB$ gives the (square of
the) Landau energy levels, with the factor $(2-\delta_{j,0})$
accounting for the  degeneracy of the $j \geq 1$ Landau
levels~\cite{Klimenko,Gusynin}.

Then, using Eq.\~(\ref{intp}) and performing the sums over the
Matsubara frequencies, the explicit expression for the effective
potential (\ref{vlde1ddim}) can be obtained. It can be written in the
form

\begin{eqnarray}
\frac{1}{N} V_{{\rm eff},\delta^1}(\sigma_c,\eta,B,T,\mu)
&=&\delta\,\frac {\sigma_c^2}{2 \lambda} +2 I_1(\eta,B,T,\mu) + 4
\delta\, \eta (\eta - \sigma_c)  I_2(\eta,B,T,\mu) \nonumber \\ &+& 2
\delta \, \frac {\lambda}{N}\left[ \eta^2 I_2^2(\eta,B,T,\mu) +
  I_3^2(\eta,B,T,\mu) \right ]\;,
\label{VeffHTmu}
\end{eqnarray}
where $I_i(\eta,B,T,\mu)$, $i=1,2,3$, are given by (see also  Appendix
\ref{definitions})

\begin{equation}
 I_1(\eta,B,T,\mu) = \frac{eB \eta}{4\pi} - \frac{(2eB)^{3/2}}{4\pi}
 \zeta \left(-\frac{1}{2},\frac{\eta^2}{2eB} \right)
 -\frac{eBT}{4\pi}\sum_{j=0}^{\infty}\alpha_j\left\{   \ln[1+
   e^{-(E_j+\mu)/T}] + \ln [1+ e^{-(E_j-\mu)/T}] \right \}\;,
\label{int1}
\end{equation}

\begin{equation}
 I_2(\eta,B,T,\mu)=- \frac{eB}{8 \pi \eta} + (2eB)^{1/2}\frac{1}{8\pi}
 \zeta\left (\frac{1}{2},\frac{\eta^2}{2eB} \right ) -
 \frac{eB}{8\pi}\sum_{j=0}^{\infty}\alpha_j \left \{\frac {1}{ E_j [1+
     e^{(E_j+\mu)/T}]} + \frac{1}{E_j  [1+ e^{(E_j-\mu)/T}]} \right
 \}\;,
\label{int2}
\end{equation}

\begin{equation}
 I_3(\eta,B,T,\mu)=\frac{eB}{8\pi}\sum_{j=0}^{\infty}\alpha_j\left[
   \frac{1}{1+e^{\left( E_j-\mu \right) /T}}-\frac{1}{1+e^{\left(
       E_j+\mu \right) /T}}\right]\,\,,
\label{int3}
\end{equation}
where $E_j=\sqrt{2jeB + \eta^2}$, $\alpha_j= 2-\delta_{j, \,0}$ and
$\zeta(s,a)$ is the Hurwitz zeta function~\cite{zeta},

\begin{equation}
\zeta(s,a) = \sum_{k=0}^{\infty} \frac{1}{(k+a)^s}\;.
\label{zetafunc}
\end{equation}

It can be easily shown from Eq.~(\ref{VeffHTmu}) that in the large-$N$
limit we reobtain the standard expression for the effective potential
for this model, as found, e.g., in the seminal papers
\cite{Klimenko,Gusynin}.
   
It is also useful to realize that Eqs.~(\ref{int2}) and  (\ref{int3})
can be both expressed in terms of Eq.~(\ref{int1}):

\begin{eqnarray}
&&I_2= -\frac{1}{2 \eta} \frac{\partial}{\partial \eta} I_1\;,
\label{altI2}
\\ &&I_3 = - \frac{1}{2} \frac{\partial}{\partial \mu} I_1\;.
\label{altI3}
\end{eqnarray}

{}Finally, let us analyze the physical meaning of the OPT $1/N$
corrections displayed by  Eq.~(\ref {VeffHTmu}). By recalling that at
one-loop order one can write the fermion number  density as

\begin{equation}
\langle \psi^+ \psi \rangle = 2 N I_3 \,\,,
\end{equation}
and the scalar condensate as

\begin{equation}
\langle {\bar \psi} \psi \rangle = - 2 N \eta I_2 \,,
\end{equation}
then, it is easy to see that Eq.~(\ref {VN1}) becomes

\begin{equation}
\Delta V_{{\rm eff},\delta^1} =  \frac {g^2}{2N} \left [ \langle
  \psi^+ \psi \rangle^2 +  \langle {\bar \psi} \psi \rangle^2 \right ]
\,,
\end{equation}
where we have used $\lambda/N\equiv g^2$. Therefore, contrary to the
LN approximation, the OPT brings in a $1/N$ suppressed term that only
contributes at finite densities.  Thus, one may expect some important
differences to arise as  this term becomes more important at
increasing $\mu$ values~\cite{ijmpe}.

As an aside, concerning the values of the coupling $\lambda$ leading to
chiral symmetry breaking, note that we can define a
renormalization condition for  the coupling as
$1/\lambda_R = 1/\lambda - 1/\lambda_c$,
with $1/\lambda_c= 4 \int d^d p/(2 \pi)^d 1/p^2$. In terms of a cutoff
regularization, in $2 < D < 4$ dimensions  $\lambda_c$ defines (for vanishing 
magnetic field) a critical value for which chiral
symmetry breaking can happen~\cite{Gusynin}, such that for
$\lambda > \lambda_c$ (i.e., corresponding to $\lambda_R <0$) the model 
can be in the broken phase of the discrete chiral symmetry, while for
$0 \leq \lambda \leq \lambda_c$ (i.e., corresponding to $\lambda_R \geq 0$), 
there is no chiral symmetry breaking. Some authors prefer to work directly 
in terms of the bare coupling $\lambda$ (like in Ref.~\cite{Gusynin}), while 
others prefer to work in terms of the redefined coupling $\lambda_R$ 
(like, for example, in Refs.~\cite{Klimenko,Vshivtsev:1996ri}).
The latter is necessarily the case when working directly in terms
of dimensional regularization, as we consider here,  since then the 
above integral vanishes by definition, and  $\lambda = \lambda_R$.
Note also that within dimensional regularization there are no additional 
divergences in the OPT case (see, e.g., Ref.~\cite{prdgn3d} for more details).

%%%%%%%%%%%%%%%%%%%%%%%%%%%%%%%%%%%%%%%%%%%%%%%%%%%%%%%%%%%%%%%%%%%%%%%

\section{Optimization and Numerical Results Beyond Large $N$}
\label{sec4}

The optimization of the effective potential, Eq. (\ref{VeffHTmu}), is
easily implemented  by applying the PMS condition, Eq. (\ref{PMS}), to
$V_{\rm eff}$.  Let us initially apply the PMS to the most general
order-$\delta$ effective potential,  which is given by Eq.~(\ref
{vlde1ddim}).  This exercise will help the reader to visualize the way
the OPT-PMS resums the perturbative series. Setting $\delta=1$ and
applying the PMS to Eq.~(\ref{vlde1ddim}), we obtain that

\begin{eqnarray}
&&\left \{ \left [ \eta - \sigma_c + \eta \frac {\lambda}{N} \left (i
    \int_p \frac {1}{P^2 - \eta^2 + i \epsilon} \right ) \right ]
  \left( 1 + \eta \frac{d}{d\eta} \right ) \left [i  \int_p \frac
    {1}{P^2 - \eta^2 + i \epsilon} \right . \right ]\nonumber \\ &
  &+\left . \frac{\lambda}{N} \left (i  \int_p \frac {P_0}{P^2 -
    \eta^2 + i \epsilon}  \right )\frac{d}{d \eta} \left (i   \int_p
  \frac {P_0}{P^2 - \eta^2 + i \epsilon} \right ) \right
  \}\Bigr|_{\eta = {\bar \eta}} =0\,\,.
\label{pmsselfconsistent}
\end{eqnarray}

{}From the result given by Eq.~(\ref{int3}), the last term of
Eq.~(\ref{pmsselfconsistent}) only survives when $\mu \ne 0$. In the
case $\mu=0$,  Eq.~(\ref{pmsselfconsistent}) factorizes in a nice way,
which allows us to understand the way the OPT-PMS procedure resums the
series producing nonperturbative results. Then, when $\mu=0$, and
using Eq. (\ref{Sigma1}), the OPT-PMS Eq.~(\ref{pmsselfconsistent})
factorizes to

\begin{equation}
\left[ {\bar \eta} - \sigma_c -  \Sigma_{\delta^1}({\bar
    \eta})|_{\mu=0} \right] \left( 1 + {\bar \eta}
\frac{\partial}{\partial{\bar \eta}} \right )  \left [i  \int_p \frac
  {1}{P^2- {\bar \eta}^2 + i \epsilon} \right ]=0\;,
\label{selfconsistent}
\end{equation}
leading to the self-consistent relation \footnote {The second solution
  is usually  discarded on the grounds that it is coupling independent
  and, moreover, does not reproduce the LN ``exact" result if one
  considers the OPT in the $N \to \infty$ limit.}

\begin{equation}
{\bar \eta} =\sigma_c + \Sigma_{\delta^1}({\bar \eta})|_{\mu=0} \;,
\label{selfcons}
\end{equation}
which is valid for any temperature and number of space-time dimensions
provided that $\mu=0$. In this way the OPT fermionic loops get
contributions containing $\sigma_c$ as well as a rainbow (exchange) type
of self-energy terms, given by Eq.~(\ref{Sigma1}).  Note that when $N
\to \infty$, ${\bar \eta}=\sigma_c$ and the large-$N$ result is
exactly reproduced~\cite{prdgn2d}. 

When $\mu \neq 0$, we can consider Eq.~(\ref{pmsselfconsistent}) in
order to get the general result in terms of the $I_i$ ($i=1,2,3$)
terms defined in the previous section.  Alternatively, using
Eqs.~(\ref{altI2}) and (\ref{altI3}) we obtain that $\bar{\eta}$ is
given by the solution of

\begin{equation}
\bar{\eta} = \sigma_c + \frac{\lambda}{2 N}  \left[ \frac{\partial
    I_1}{\partial \eta} + \frac{ \left(\frac{\partial I_1}{\partial
      \mu}\right) \left(\frac{\partial^2 I_1}{\partial \eta \partial
      \mu} \right)}{ \frac{\partial^2 I_1}{\partial \eta^2} }
  \right]\Bigr|_{\eta = {\bar \eta}}\;.
\label{etabar}
\end{equation}
Again, we see from Eq.~(\ref{etabar}) that as $N\to \infty$, the
large-$N$ result $\bar{\eta}=\sigma_c$ is recovered.

The PMS equation (\ref{etabar}) is to be solved together with the one
defining the vacuum expectation value for the background field
$\sigma_c$,

\begin{eqnarray}
\frac{ \partial V_{\rm eff}}{\partial \sigma_c} \Bigr|_{\sigma_c={\bar
    \sigma}, \eta =\bar \eta} =0\,\;\;\; \Rightarrow  \bar\sigma=  4
\lambda \, \bar \eta \, I_2\,.
\label{barsigma}
\end{eqnarray}
We now have all the necessary tools to investigate all the possible
cases when considering a finite constant magnetic field  applied to
the system. As a ballpark estimate we shall consider magnetic fields
ranging from $eB=0$ to $eB=20\, \Lambda^2$, since the gap energy is
about  $\Lambda$ within this model. This choice is reasonable since,
keeping in mind possible applications to condensed matter systems for
example, typically, the gap energy lies within the range $10-100 \;
{\rm meV}$  and if,  for example, one considers the lower gap value
$\Lambda \sim 10 \; {\rm meV}$ then  $eB=20 \Lambda^2 \sim 3$ Teslas,
which is a realistic value within current planar condensed matter
systems\footnote{When converting our results to condensed matter
  systems, note that  one should also include explicitly the Fermi
  velocity in the expresssions~\cite {poly}.}.  Finally, we set $N=2$,
since this is the relevant value as far as  planar condensed matter
systems (like high-temperature superconductor films or graphene) are
concerned.

%%%%%%%%%%%%%%%%%%%%%%%%%%%%%%%%%%%%%%%%%%%%%%%%%%%%%%%%%%%%%%%%%%%

\subsection{$T= 0$ and $ \mu = 0$ case}

Let us preliminarily examine the case of zero temperature and zero
chemical potential, $T=\mu=0$. An important effect here is that of
magnetic catalysis~\cite{Klimenko,Gusynin}, which  we next investigate
in order to analyze how this phenomenon is affected by the
nonperturbative inclusion of finite $N$ corrections through the OPT.
{}From Eq.~(\ref{VeffHTmu}) for the effective potential and upon using
the results (\ref{I1Tmu0}), (\ref{I2Tmu0}), and (\ref{I3Tmu0})
obtained in the  appendix, we have that

\begin{eqnarray}
\frac{1}{N} V_{{\rm eff},\delta^1}(\sigma_c,\eta,B,T=0,\mu=0) &= &
\delta \frac {\sigma_c^2}{2 \lambda} +\frac{eB \eta}{2\pi} -
\frac{(2eB)^{3/2}}{2\pi} \zeta\left (-\frac{1}{2},\frac{\eta^2}{2eB}
\right ) \nonumber \\ &-& \delta\frac {\eta(\eta-\sigma_c)}{2\pi}\left
       [ \frac{eB}{ \eta} -  (2eB)^{1/2} \zeta\left
         (\frac{1}{2},\frac{\eta^2}{2eB} \right ) \right ] \nonumber
       \\ &+&\delta \frac{ \lambda \eta^2}{32\pi^2 N}\left [
         \frac{eB}{ \eta} -  (2eB)^{1/2} \zeta\left
         (\frac{1}{2},\frac{\eta^2}{2eB} \right ) \right ]^2\;.
\label{VeffH}
\end{eqnarray}
Then, from Eq.~(\ref{etabar}), we  obtain the self-consistent relation
to be evaluated for $\bar{\eta}$:

\begin{equation}
 {\bar \eta}=\sigma_c + \frac{ \lambda}{8\pi N}\left [ eB -   {\bar
     \eta} \, (2eB)^{1/2}  \zeta\left (\frac{1}{2},\frac{{\bar
       \eta}^2}{2eB} \right ) \right ] \,.
\label{etabarh}
\end{equation}

\begin{figure}[htb]
  \vspace{0.5cm}  \epsfig{figure=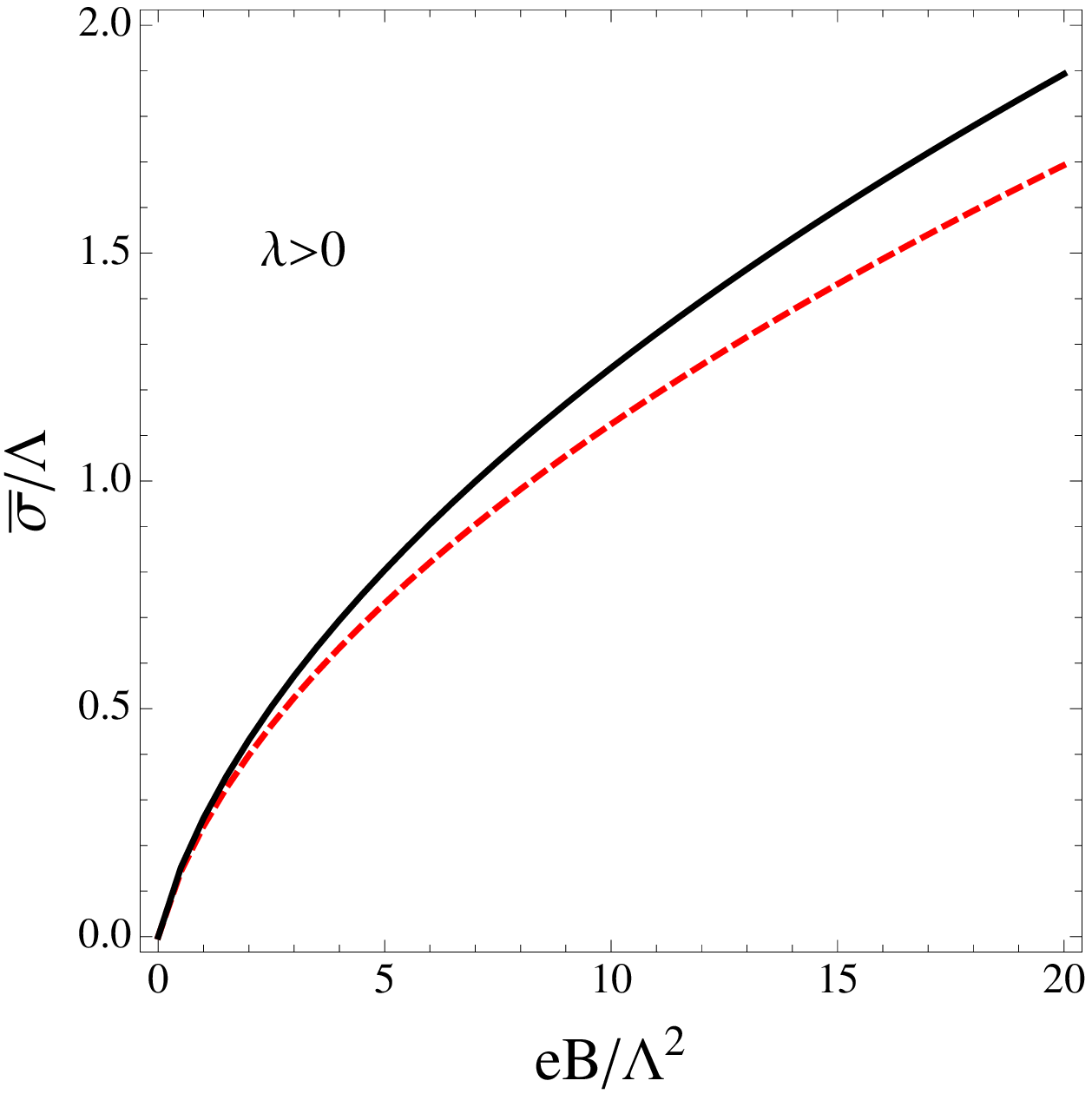,angle=0,width=7cm}
  \epsfig{figure=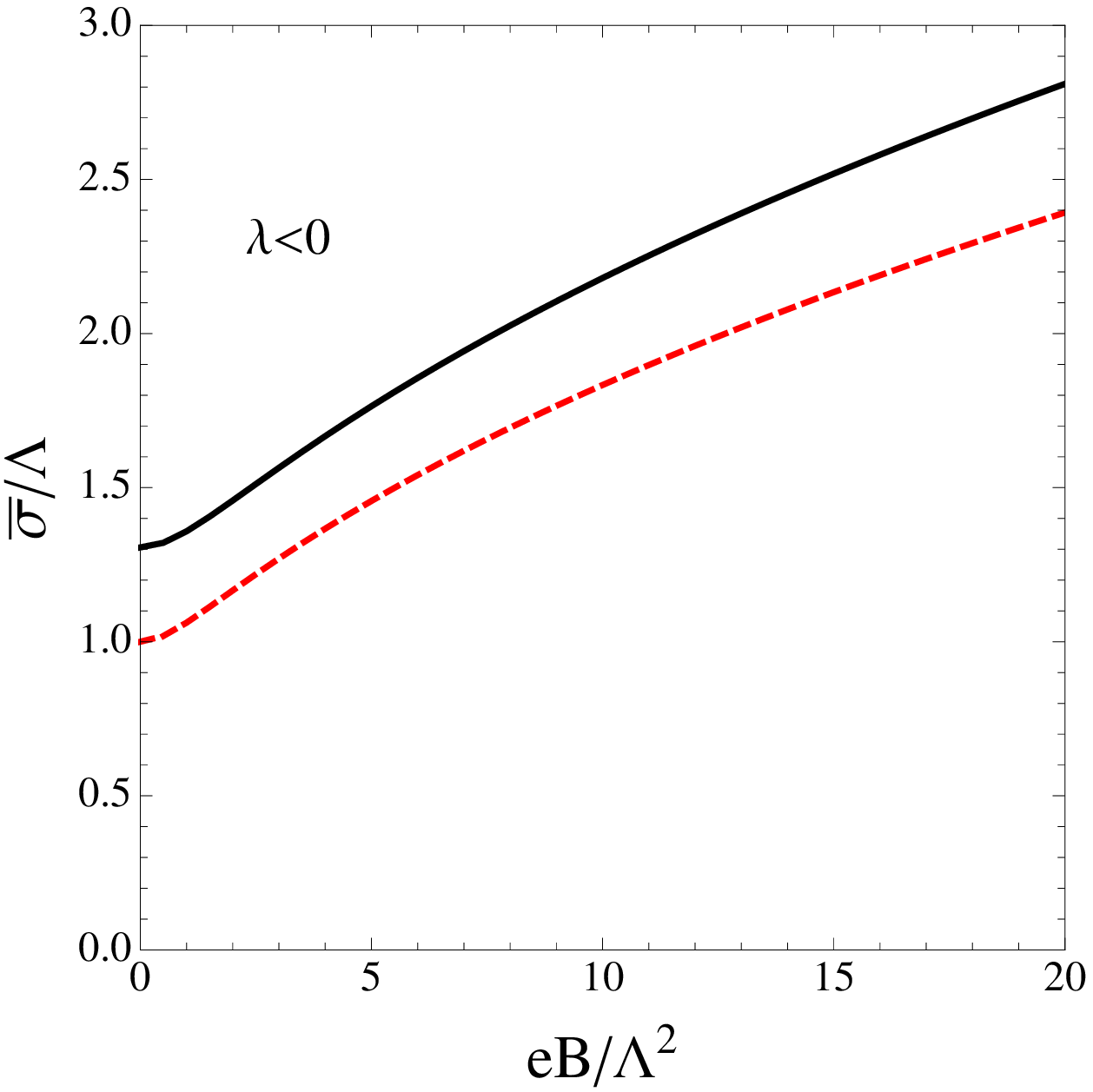,angle=0,width=7cm}
\caption[]{The order parameter $\bar \sigma$, as a function of $eB$
  and for $N=2$, illustrating the phenomenon of magnetic catalysis at
  $T=0$ and $\mu=0$.  The dashed line is the large $N$ result, and the
  continuous line represents the OPT result. The left panel shows the
  case $\lambda >0$, illustrating how a finite magnetic field induces
  CSB. The right panel shows the situation for the $\lambda < 0$ case,
  when CSB occurs even at $B=0$. }
\label{sigmaH}
\end{figure}

By extremizing the effective potential (\ref{VeffH}) with respect to
$\sigma_c$, setting $\delta=1$ and using the optimal ${\bar \eta}$,
one finds that the order parameter satisfies

\begin{equation}
 \frac{\bar \sigma}{\lambda} = - \frac{1}{2\pi} \left [ eB -   {\bar
     \eta} \, (2eB)^{1/2} \zeta\left ( \frac{1}{2},\frac{{\bar
       \eta}^2}{2eB} \right ) \right ] \,.
\label{barsigmaH}
\end{equation}
Then,  comparing the above equation with Eq.~(\ref{etabarh}), one is
lead to the relation

\begin{equation}
 {\bar \eta}= {\bar \sigma} \, {\cal F}(N) \,\,,
\label{etaF}
\end{equation}
where

\begin{equation}
{\cal F}(N) = 1 - \frac {1}{4N} \;.
\label{FN}
\end{equation}

By using Eq.~(\ref{etaF}) in Eq.~(\ref{barsigmaH}), we obtain

\begin{equation}
 \frac{\bar \sigma}{\lambda} = - \frac{eB}{2\pi} + \frac{{\bar
     \sigma}{\cal F}(N)}{2\pi}  (2eB)^{1/2} \zeta\left (
 \frac{1}{2},\frac{{{\bar \sigma}^2{\cal F}(N)}^2}{2eB} \right )  \,.
\label{barsigmaHsol}
\end{equation}

Chiral symmetry breaking can now be investigated by looking at the
case $\lambda > 0$, in which CSB does not occur when $B=0$, and also
the case $\lambda <0$, where   dynamical CSB occurs even at $B=0$. All
quantities will be expressed in terms of the scale
$\Lambda=\pi/|\lambda|$, which is the value of the chiral condensate
$\bar \sigma$ in the LN case.  Our numerical results are compared to
the ones given by the LN approximation in {}Fig.~\ref{sigmaH} for
$N=2$. We note that when either $\lambda >0$ or $\lambda <0$,  the
magnetic catalysis is enhanced by the finite $N$ contributions that
the OPT-PMS method brings, as  {}Fig \ref{sigmaH} shows.   

%%%%%%%%%%%%%%%%%%%%%%%%%%%%%%%%%%%%%%%%%%%%%%%%%%%%%%%%%%%%%%%%%%%%%%%%

\subsection{$T \ne 0$ and $ \mu = 0$ case}

{}As in the previous subsection, we start with Eq.~(\ref{VeffHTmu})
for the free energy density and use the results (\ref{I1mu0}),
(\ref{I2mu0}) and (\ref{I3mu0})  to write

\begin{eqnarray}
\frac{1}{N} V_{{\rm eff},\delta^1}(\sigma_c,\eta,B,T,\mu=0)
&=&\delta\,\frac {\sigma_c^2}{2 \lambda} +2  \left[ \frac{eB
    \eta}{4\pi} - \frac{(2eB)^{3/2}}{4\pi} \zeta
  \left(-\frac{1}{2},\frac{\eta^2}{2eB} \right)
  -\frac{eBT}{2\pi}\sum_{j=0}^{\infty}\alpha_j \ln\left(1+
  e^{-E_j/T}\right) \right] \nonumber \\ &+& 4 \delta\, \eta (\eta -
\sigma_c) \left[ - \frac{eB}{8 \pi \eta} +  \frac{(2eB)^{1/2}}{8\pi}
  \zeta\left (\frac{1}{2},\frac{\eta^2}{2eB} \right ) -
  \frac{eB}{4\pi}\sum_{j=0}^{\infty}\alpha_j \frac{1}{ E_j \left(1+
    e^{E_j/T} \right) } \right] \nonumber \\ &+& 2 \delta \, \frac
      {\lambda}{N} \eta^2 \left[  - \frac{eB}{8 \pi \eta} +
        \frac{(2eB)^{1/2}}{8\pi} \zeta\left
        (\frac{1}{2},\frac{\eta^2}{2eB} \right ) -
        \frac{eB}{4\pi}\sum_{j=0}^{\infty}\alpha_j \frac{1}{ E_j [1+
            e^{E_j/T}]} \right]^2 \;,
\label{VeffHTmu0}
\end{eqnarray}

The expression for the order parameter ${\bar \sigma}$, equivalent to
Eq.~(\ref{barsigmaHsol}) in the case of $T=0$, now reads

\begin{equation}
 \frac{\bar \sigma}{\lambda} = - \frac{eB}{2\pi} + \frac{{\bar
     \sigma}{\cal F}(N)}{2\pi}  (2eB)^{1/2} \zeta\left (
 \frac{1}{2},\frac{{{\bar \sigma}^2{\cal F}(N)}^2}{2eB} \right )
 -\frac{eB}{4\pi} \sum_{j=0}^{\infty}\alpha_j \frac {1}{
   E_j(\bar{\sigma}) [1+ e^{E_j(\bar{\sigma})/T}]}\;,
\end{equation}
where $ E_j(\bar{\sigma}) = \sqrt {2jeB + {\bar \sigma}^2 {\cal
    F}^2(N)}$ and we have used  Eq.~(\ref{etaF}), which still holds at
$T \ne 0$ and $ \mu = 0$.

\begin{figure}[htb]
  \vspace{0.5cm}  \epsfig{figure=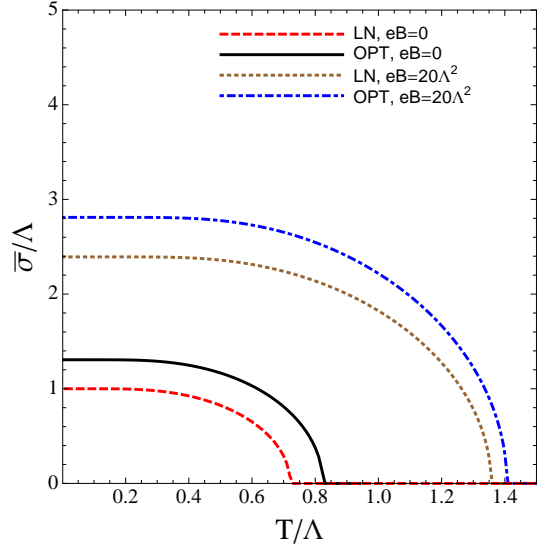,angle=0,width=7cm}
\caption[]{The order parameter, $\bar \sigma/\Lambda$, as a function
  of $T/\Lambda$ for $N=2$ at $\mu=0$ for $B=0$ and $eB= 20\,
  \Lambda^2$. The dashed lines represent the large-$N$ result and the
  continuous lines represents the OPT result.  The figure illustrates
  a transition of the second kind for all the four cases.}
\label{sigT}
\end{figure}

\begin{figure}[htb]
  \vspace{0.5cm}  \epsfig{figure=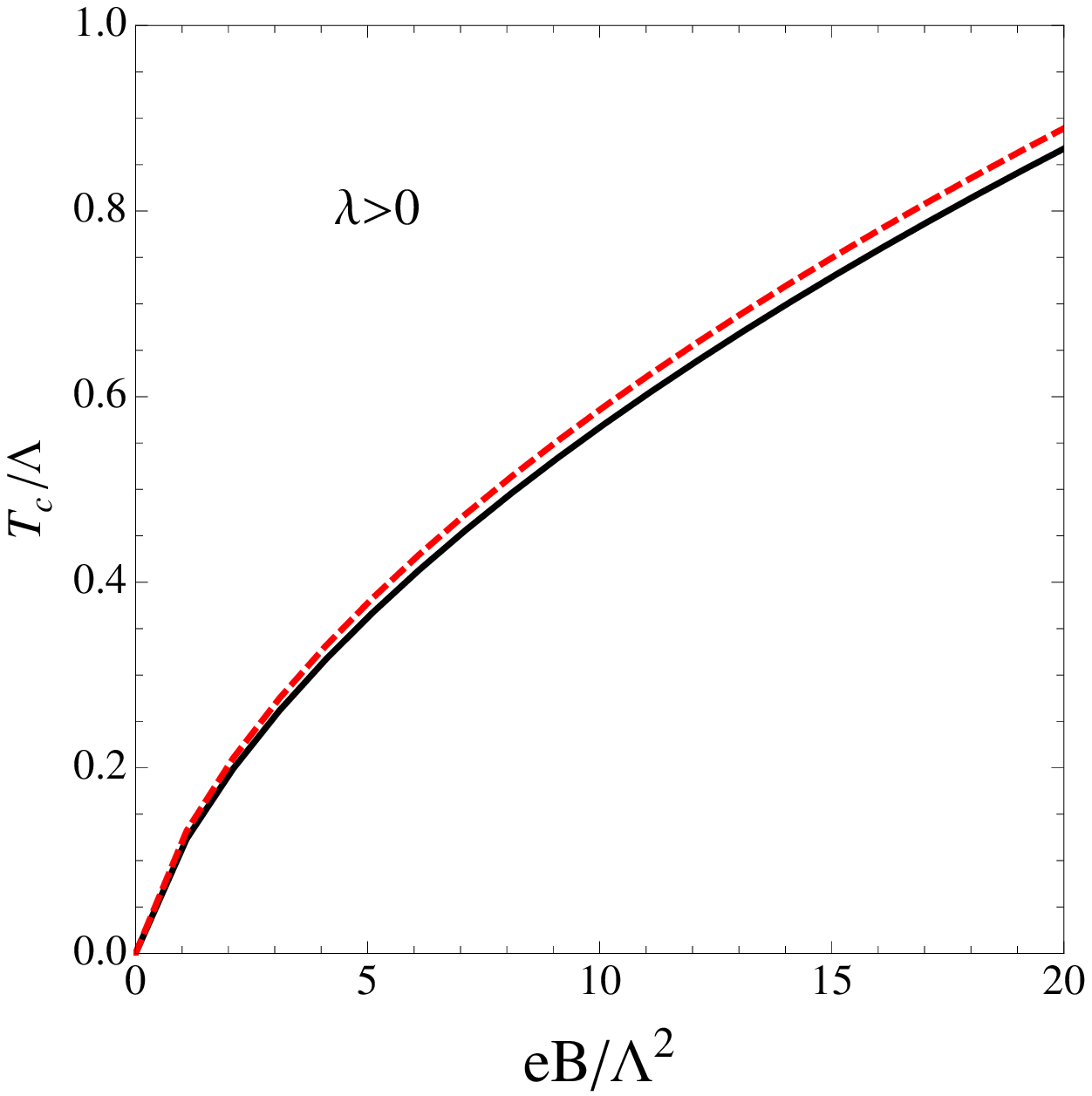,angle=0,width=7cm}
  \epsfig{figure=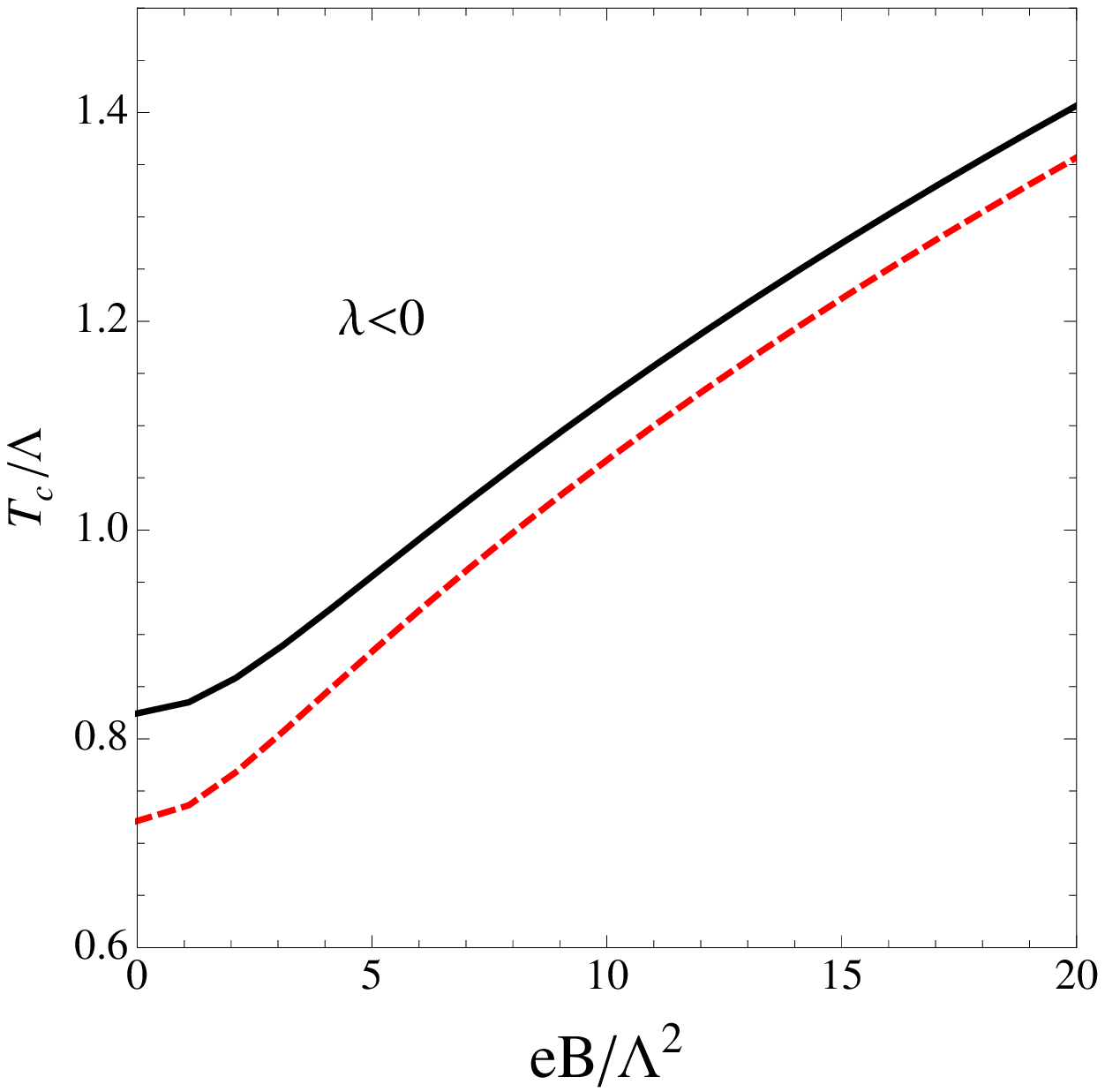,angle=0,width=7cm}
\caption[]{The critical temperature, $T_c/\Lambda$,  as a function of
  $eB/\Lambda^2$ for $N=2$ and $\mu=0$. The dashed lines represent the
  large-$N$ result and the continuous lines represent the OPT
  result. The left panel is for $\lambda > 0$ and the right panel for
  $\lambda < 0$. }
\label{TcH}
\end{figure}

The thermal behavior for the order parameter, ${\bar \sigma}(T)$, is
shown in {}Fig.~\ref{sigT} for $eB=0$ and for $eB= 20 \,
\Lambda^2$. In both cases the transition is of the second kind and, as
expected, in the later case the symmetry restoration happens at a
higher $T_c$. {}Figure \ref{TcH} shows how the critical temperature
increases with $B$, which is expected since $\bar \sigma$ increases
with $B$ and $T_c \sim {\bar \sigma}$. In this model even a strong
magnetic field (e.g., $eB \sim 30 \Lambda^2$) is not able to change
the character of the phase transition.  It is interesting to note that
for $\lambda >0$ the OPT predicts that the order parameter assumes
higher values than the ones predicted by the LN approximation   as $B$
increases as the left panel of Fig.~\ref{sigmaH} suggests. However,
despite the fact  that chiral symmetry seems to be more severely broken within the
OPT framework, the left panel of {}Fig.~\ref{TcH}  shows that this
symmetry will be restored at smaller critical temperatures than  those
predicted by the LN approximation. 

%%%%%%%%%%%%%%%%%%%%%%%%%%%%%%%%%%%%%%%%%%%%%%%%%%%%%%%%%%%%%%%%%%%%%%%%%%%%%

\subsection{$T = 0$ and $ \mu \ne 0$ case}
\label{T0section}

The next case we analyze is when $T = 0$ and $ \mu \ne 0$, which is
relevant, for example, when analyzing charge
asymmetries~\cite{chargeasy}.  The OPT first-order result for the
effective potential in this case  becomes

\begin{eqnarray}
\frac{1}{N} V_{{\rm eff},\delta^1}(\sigma_c,\eta,B,T=0,\mu)
&=&\delta\,\frac {\sigma_c^2}{2 \lambda} +2 I_1(\eta,B,T=0,\mu) + 4
\delta\, \eta (\eta - \sigma_c)  I_2(\eta,B,T=0,\mu) \nonumber \\ &+&
2 \delta \, \frac {\lambda}{N}\left[ \eta^2 I_2^2(\eta,B,T=0,\mu) +
  I_3^2(\eta,B,T=0,\mu) \right ]\;,
\label{VeffHT0mu}
\end{eqnarray}
where the expressions for $I_i(\eta,B,T=0,\mu)$, $i=1,2,3$, are again
as those  given in the Appendix \ref{definitions} by
Eqs.~(\ref{I1T0}), (\ref{I2T0}) and (\ref{I3T0}), respectively.

Then, using Eq.~(\ref{barsigma}) one can write the order parameter as

\begin{equation}
\bar \sigma \equiv 4 \lambda I_2(\eta,B,T=0,\mu) = 4 \lambda \bar \eta
\left[  -\frac{eB}{8 \pi \eta} + \frac{(2eB)^{1/2}}{8\pi} \zeta\left
  (\frac{1}{2},\frac{\eta^2}{2eB} \right)- \frac{eB}{8 \pi}
  \sum_{j=0}^{\frac{\mu^2 - \eta^2}{2 eB}} \alpha_j \frac{1}{E_j}
  \theta(\mu-\eta)\right]\;,
\label{barsigmaT0mu}
\end{equation}
while the PMS equation  for $\mu\ne 0$ now takes the more general form
of Eq.~(\ref{etabar}), which after some algebra can be expressed for
$T=0$ as 

\begin{equation}
\bar \eta = \bar \sigma {\cal F}(N) + \frac{\lambda}{4\pi\:N} \eta
\left[ \frac{I_3(\eta,B,T=0,\mu)}{(1+\eta
    \frac{\partial}{\partial\eta})I_2(\eta,B,T=0,\mu)}\right]\;,
\label{baretaT0mu}
\end{equation}
where the relevant expressions for $I_i$ at $T=0$ are given in
Appendix \ref{definitions} in Eqs.~(\ref{I2T0})-(\ref{I3T0})  and the
OPT corrections for $\mu\ne 0$ to the simpler relation in
Eq.~(\ref{etaF}), $ {\bar \eta}= {\bar \sigma} \, {\cal F}(N) $, are
explicit from the second term. In Eq.~(\ref{baretaT0mu}) we have used 

\begin{equation}
\frac{\partial}{\partial\eta} I_3(\eta,B,T,\mu) |_{T\to 0} =
-\frac{\eta}{4\pi}
\label{dI3eta}
\end{equation}
which is independent of $B\;$~\footnote{Caution should be taken to take
  the derivative with respect to $\eta$ of $I_3(\eta,B,T,\mu)$ before
  taking the $T\to 0$ limit; otherwise, the derivative of expression
  (\ref{I3T0}) is ill-defined. The result (\ref{dI3eta}) is
  consistent with the $B\to 0$ limit~\cite{prdgn3d}  for $\mu\ne 0$ of
  expression (\ref{baretaT0mu})}. Note, however, that the second term in
(\ref{baretaT0mu}) being suppressed by $(4\pi N)^{-1}$ gives a
reasonably small correction, moreover only nonvanishing for $\mu
>\eta$ due to the step function in $I_3$ Eq.~(\ref{I3T0}). A
legitimate  approximation can thus be to use the simpler relation
Eq.~(\ref{etaF}) {\em within} this correction, instead of the implicit
exact $\eta$  relation in (\ref{baretaT0mu}), since the difference is
of higher $\lambda$ order, neglected anyway at the first OPT
$\delta$-order here considered.  It is worth remarking that this OPT
correction term, when nonvanishing, may be positive or negative
depending on the sign of $\lambda$ and depending on the sign of
$I_2(\eta,B,T=0,\mu)$ (while $I_3(\eta,B,T=0,\mu)>0$ for any $B$
values).  Thus, it may enhance $\eta$ with respect to the LN  result
$\eta=\sigma_c$ (partly compensating the reduction from ${\cal F}(N) <
1$). Using  the $T=0$ analytical expression of $I_2$ and $I_3$ in
Eqs.~(\ref{I2T0}) and (\ref{I3T0}) and some properties of the
Riemann--Hurwitz Zeta functions, it is not difficult to recover the
$B\to 0$ limit of (\ref{baretaT0mu}), having relatively simple
expressions:
 
\begin{equation}
\bar \eta(T=0,B\to 0) = \bar \sigma {\cal F}(N)
-\frac{\lambda}{8\pi\:N} \frac{\eta
  (\mu^2-\eta^2)\theta(\mu-\eta)}{\eta +(\mu-\eta)\theta(\mu-\eta)}\;,
\end{equation}
which is consistent with the direct $B=0$ calculation~\cite{prdgn3d}.
\\ Here, as we shall see, chiral symmetry is restored through a
first-order phase transition as in the case of the absence of the
external magnetic field~\cite{prdgn3d}. Therefore, we must determine
the coexistence chemical potential value,  $\mu_c$, at which the
discontinuous chiral symmetry transition occurs.  In this case,
$\mu_c$ is obtained by  solving (see also discussion in the next
subsection)

\begin{equation}
V_{\rm eff}(\sigma_c=\bar{\sigma},B,T=0,\mu_c) = V_{\rm
  eff}(\sigma_c=0,B,T=0,\mu_c)\;.
\label{1stlineT0}
\end{equation}

\begin{figure}[htb]
  \vspace{0.5cm}  \epsfig{figure=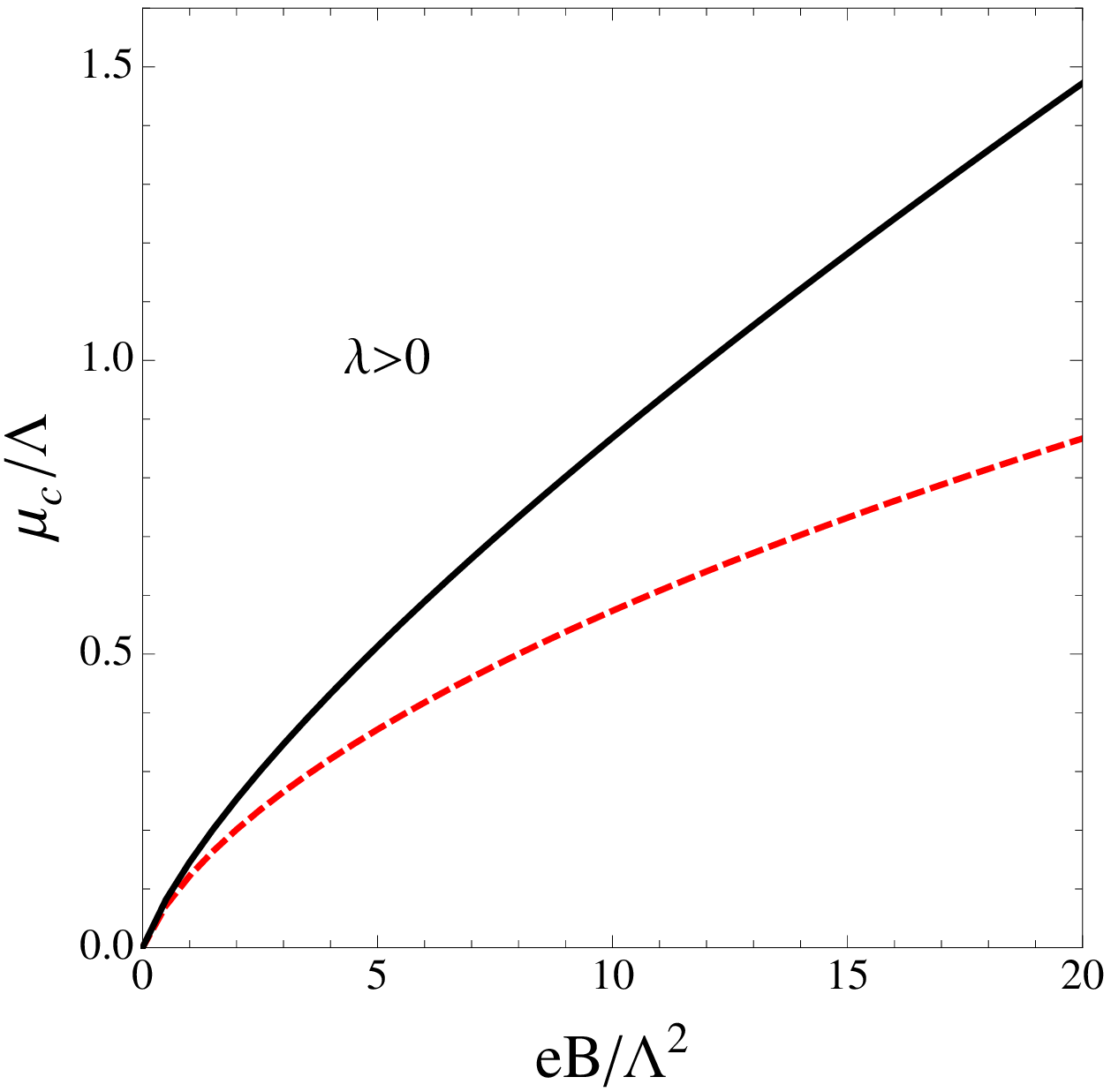,angle=0,width=7cm}
  \epsfig{figure=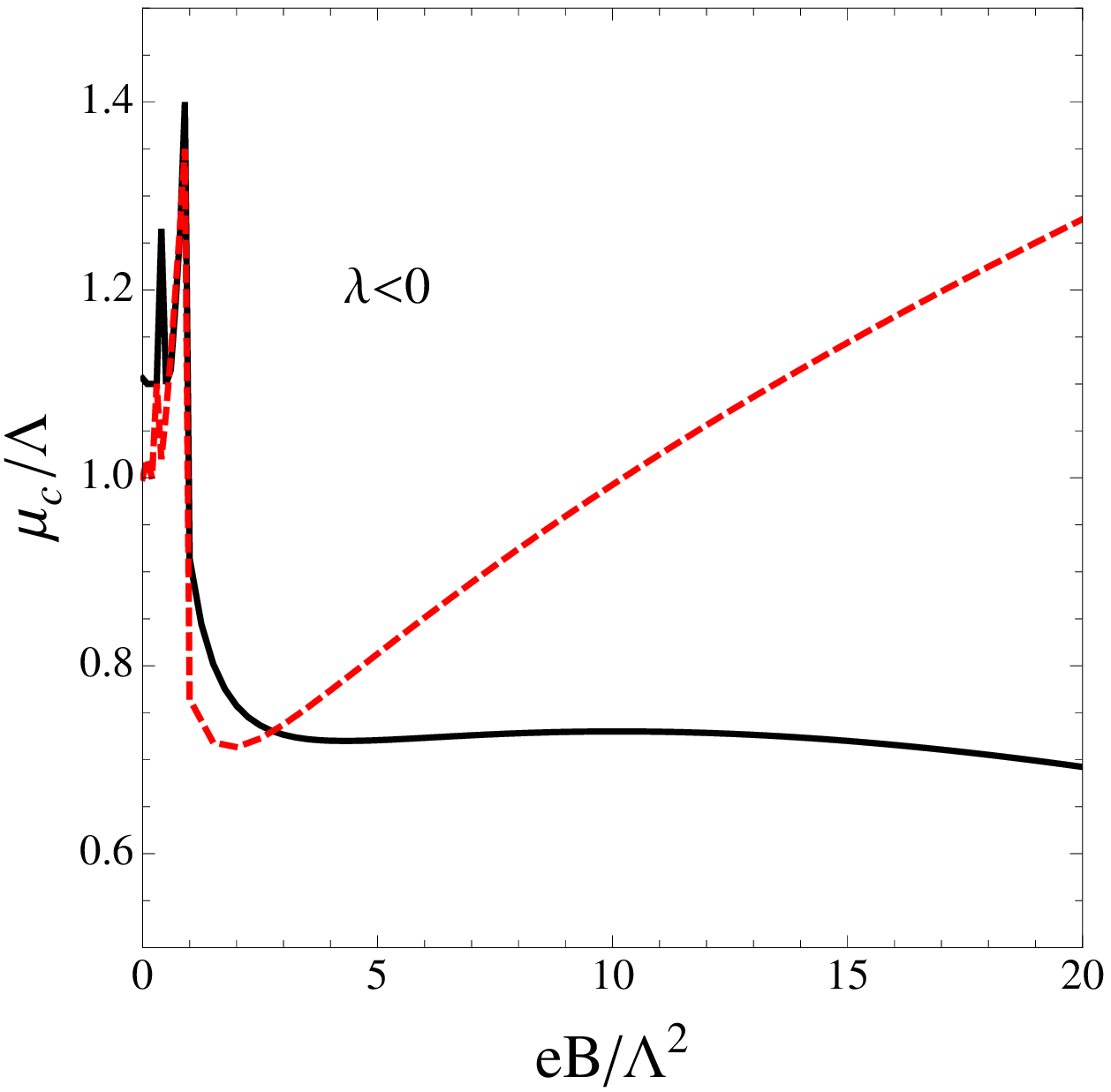,angle=0,width=7cm}
\caption[]{The coexistence chemical potential in units of $\Lambda$ as
  a function of $eB/\Lambda^2$ for $T=0$. The left panel illustrates
  the $\lambda >0$ case and the right panel the $\lambda < 0$
  case. The OPT predicts that IMC takes
  place for $eB \gtrsim \Lambda^2$ when $\lambda < 0$ contrary to the
  large-$N$ result which  displays this phenomenon only at for $
  \Lambda^2 \lesssim eB \lesssim 2 \Lambda^2$. }
\label{figmuc}
\end{figure}

\begin{figure}[htb]
  \vspace{0.5cm}  \epsfig{figure=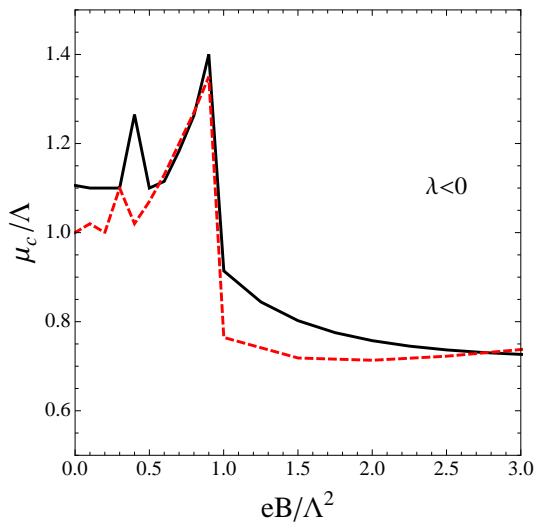,angle=0,width=7cm}
\caption[]{de Haas--van Alphen oscillations observed when the
  coexistence chemical potential, $\mu_c/\Lambda$, varies within the
  $0< eB \lesssim \Lambda^2$ range at $T=0$ when $\lambda <0$.}
\label{figmuclowH}
\end{figure}

Results for $\mu_c$ as a function of the magnetic field, for both
cases of $\lambda<0$ and $\lambda>0$, in the LN and OPT cases, are
shown in {}Fig.~\ref{figmuc}. For $eB \le  \, \Lambda^2$  and $\lambda
<0$, one observes the typical de Haas--van Alphen oscillations (see
Appendix \ref{oscillations})  due to the filling of the Landau levels.
In {}Fig.~\ref{figmuclowH} we show the critical chemical potential
$\mu_c(B)$ for $\lambda<0$ for the region of low magnetic fields. This
figure shows more clearly the typical oscillations at low magnetic
fields, which are reminiscent  of the de Haas--van Alphen magnetic
oscillations of the magnetization.  These oscillations stop after $eB$
reaches a value such that only the lowest Landau level (LLL) has to be
considered (here this happens at $eB \gtrsim \, \Lambda^2$). After
this point one sees a remarkable difference between the OPT and the LN
approximation results. The latter predicts that $\mu_c$ decreases with
$eB$ toward a minimum  and then observes a sharp increase for $eB
\gtrsim 2\, \Lambda^2$. This LN result is in complete agreement with a
MFA application to the three-flavor NJL model  performed in
Ref.~\cite{prc} (see Ref.~\cite{andre} for a detailed discussion on
the  first-order coexistence region). In contrast the decrease of
$\mu_c$ with $B$ in the OPT case is a manifestation of the inverse
magnetic catalysis (IMC) effect, which was explained e.g. in
Ref.~\cite{andreas} (the IMC effect was first observed in the NJL
model in  Ref.~\cite{IMCklimenko} at $T = 0$ and in Ref.~\cite{IMC2} for the
full  $T-\mu-B$ case).  Then, {}Fig.~\ref{figmuc} shows that the OPT
results is more in line with this phenomenon, since only a smooth very
moderate rise of $\mu_c$ is observed to occur between $eB \simeq 5\,
\Lambda^2$ and $eB \simeq 15\, \Lambda^2$,  before it drops again at
higher fields. This quantitative difference can be traced to the  OPT
$\lambda\langle \psi^\dagger \psi\rangle/N$ type of corrections,
which are non-negligible in this region of high charge asymmetry. In
fact the behavior can be essentially understood from a simple
analytical approximation.  {}First note that both terms of
Eq.~(\ref{1stlineT0}) considerably simplify. Because of $\sigma_c=0$
the right-hand side can be written as

\begin{equation}
 V_{\rm  eff}(\sigma_c=0,B,T=0,\mu) = 2I_1(0,B,0,\mu)
 +2\frac{\lambda}{N} I^2_3(0,B,0,\mu)\;.
\label{Vsig0T0}
\end{equation}
Also, calculating $V_{\rm eff}$ from Eq.~(\ref{VeffHT0mu}) at its
minimum, using   the relation Eq.~(\ref{barsigma}) between
$\bar\sigma_c$ and $\eta$ (which is valid at the minimum of the
potential for any values of the other parameters), the left-hand-side
of Eq.~(\ref{1stlineT0}) also simplifies to
 
\begin{equation}
 V_{\rm  eff}(\sigma_c,B,T=0,\mu) = \frac{\bar\sigma^2_c}{2\lambda}
 {\cal F}(N)+2I_1(\bar\sigma_c,B,0,\mu) +2\frac{\lambda}{N}
 I^2_3(\bar\sigma_c,B,0,\mu)\;.
\label{VsigT0}
\end{equation}
Moreover, the last term in Eq.~(\ref{VsigT0}) vanishes whenever
$\bar\eta >\mu$, which is the case in most of the parameter space
considered, i.e. $\mu_c$ satisfying Eq.~(\ref{1stlineT0}) will be such
that $\mu_c < \bar \eta$.  The LN case can be easily recovered from
the above expressions by simply neglecting the OPT correction $\lambda
I^2_3$ term, and taking ${\cal F}(N)=1$, $\bar\eta=\sigma_c$ in the
remaining terms.  {}For $e B \gtrsim \, \Lambda^2$ only the lowest
Landau level contributes to the relevant integrals, such that in this
range $e B \gtrsim \, \Lambda^2$ Eq.~(\ref{1stlineT0}) gives a
relatively simple analytic (implicit) expression for $\mu_c$ in the
OPT case:

\begin{equation}
\mu^{\rm OPT}_c = - {\cal F}(N)\bar\sigma_c +{\cal F}(N)
\frac{\bar\sigma^2_c}{e B}+  2(2 e
B)^{1/2}\left[\zeta\left(-1/2,\frac{{\cal F}^2(N)\bar\sigma^2_c}{2 e
    B}\right)- \zeta(1/2)\right] +\frac{\lambda}{16\pi N\,e B}\left(e
B+\mu^2_c \right)^2\;,
\label{mucOPT}
\end{equation}
while the corresponding expression in the LN case reads

\begin{equation}
\mu^{\rm LN}_c = -\bar\sigma_c + \frac{\bar\sigma^2_c}{e B} + 2(2 e
B)^{1/2}  \left[ \zeta\left(-1/2,\frac{\bar\sigma^2_c}{2 e
    B}\right)-\zeta(1/2) \right]\;.
\label{mucLN}
\end{equation}
The exact $e B$ dependence in Eqs.~(\ref{mucOPT}) and (\ref{mucLN}) is
rather involved, since  $\sigma_c$ depends nontrivially on $e B$,
Eq.~(\ref{barsigmaT0mu}).  However, for a qualitative but essentially
rather accurate understanding of the behavior in  {}Fig.~\ref{figmuc},
it is sufficient to know that  $\sigma_c(e B)$ is a moderately increasing
function of $e B$. Then the last term in Eq.~(\ref{mucLN}) involving
the $\zeta$-functions is monotonically increasing with $e B$ (first rapidly
for moderate $e B$ and then for large $e B$ with a decreasing slope),
so that together with the first terms it implies    that $\mu^{\rm
  LN}_c(e B)$ gets a minimum at a moderate $e B$ value, and then has a
steeper rise. Now, if there would only be the moderate  difference
$\sigma_c \to \sigma_c {\cal F}(N)$ from the LN to the OPT case, the
OPT results would be qualitatively similar to the LN ones.  In
contrast, due to the last correction  term in Eq.~(\ref{mucOPT}), the
behavior of $\mu^{\rm OPT}_c( e B)$ is drastically different, since
for $\lambda <0$ the last terms goes for large $e B$ as $-(e B)/(16N)$
(the $\mu^2$ in the last term being rapidly negligible in the relevant
range $e B \gg \mu^2$), which thus prevents $\mu^{\rm OPT}_c$ to
increase fast, producing almost a plateau, before this term starts to
drive $\mu_c$ to decrease for even larger values of $e B$. 
Clearly the opposite behavior happens for $\lambda >0$, as seen
in {}Fig.~\ref{figmuc} (left panel).  Of course,
for extremely large $e B$ values the OPT correction term will become
an unreasonably large perturbative correction and not very trustable,
since higher $\lambda$-order corrections are not considered at the OPT
first order.

%%%%%%%%%%%%%%%%%%%%%%%%%%%%%%%%%%%%%%%%%%%%%%%%%%%%%%%%%%%%%%%%%%%%%%%%%

\subsubsection{Intermediate transitions at low magnetic field}

\begin{figure}[htb]
  \vspace{0.5cm}  \epsfig{figure=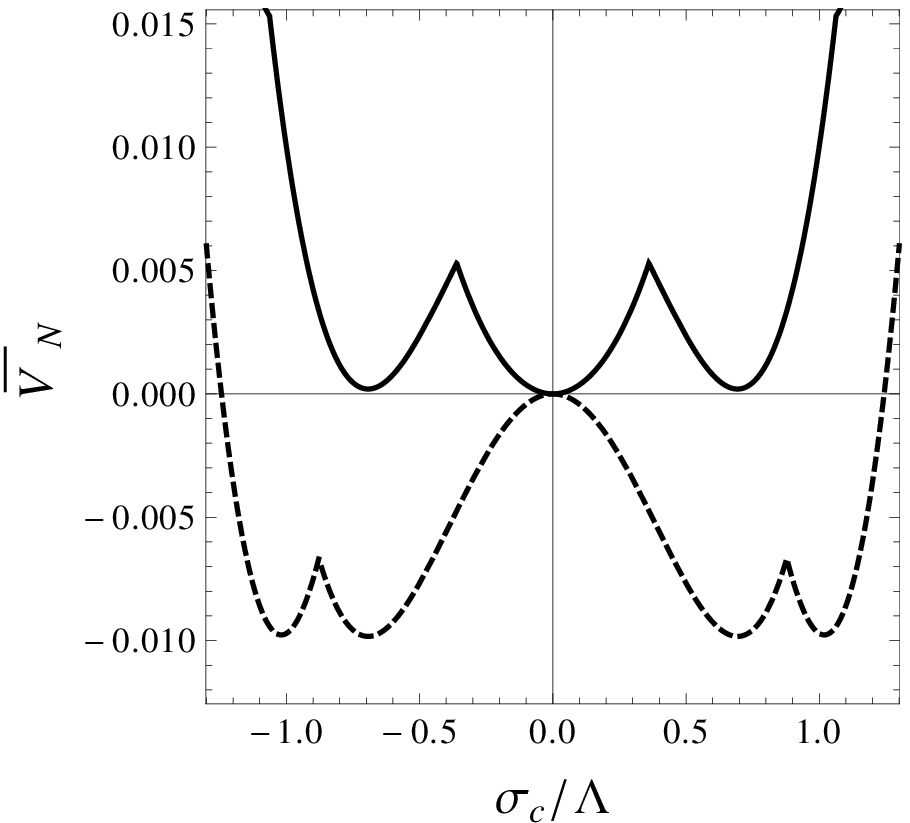,angle=0,width=7.5cm}
\hspace{0.25cm} \epsfig{figure=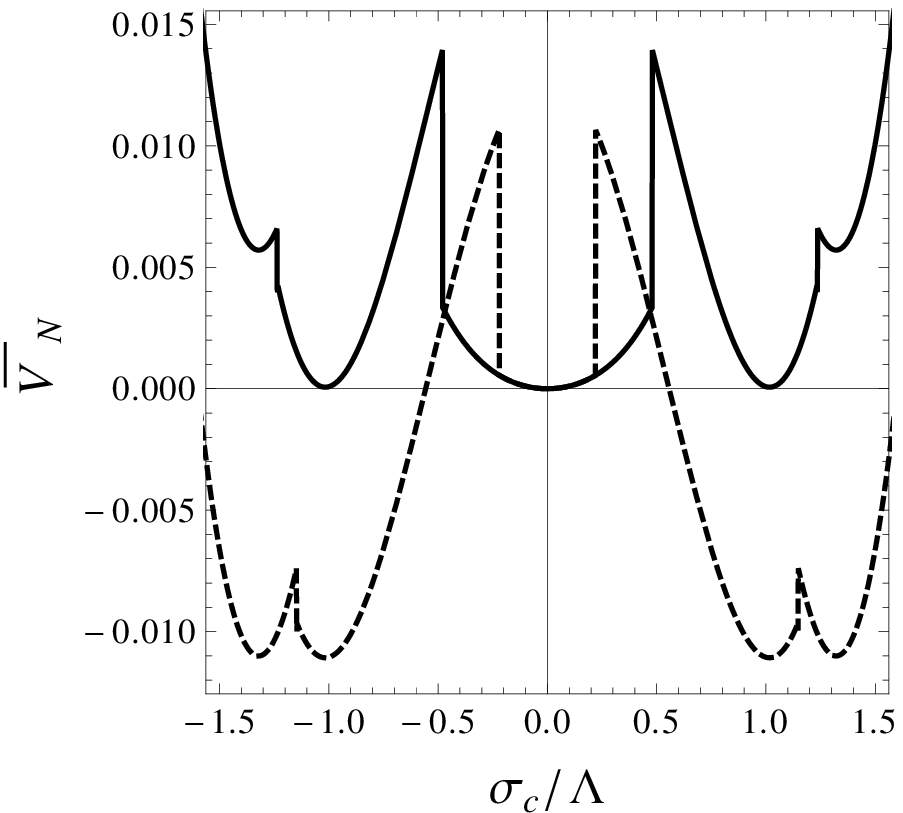,angle=0,width=7.5cm}
\caption[]{The (normalized) effective potential (for $\lambda <0$)  at
  $T=0$ and $eB=0.5 \Lambda^2$ for the LN case (the plot on the left)
  and for the OPT (the plot on the right) with $N=2$.  The dashed line
  is for $\mu \simeq 0.878 \Lambda$ in LN case, while in the OPT it is
  $\mu \simeq 1.02 \Lambda$.  The solid line is for $\mu \simeq 1.063
  \Lambda$ in LN case, while in the OPT it is $\mu \simeq 1.09
  \Lambda$.  }
\label{multiplePT}
\end{figure}

\begin{figure}[htb]
  \vspace{0.5cm}  \epsfig{figure=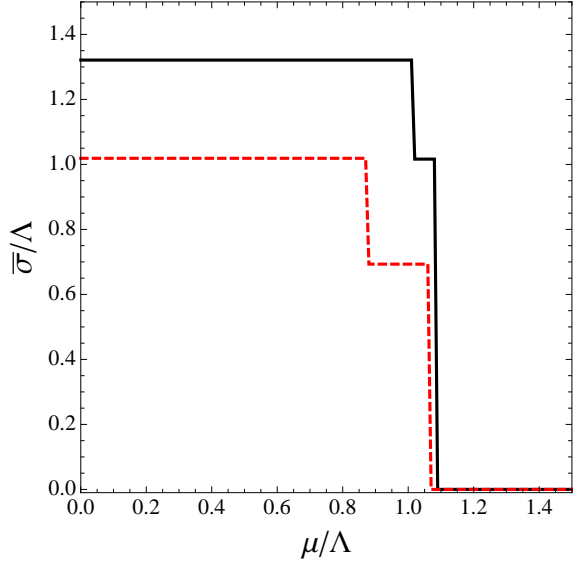,angle=0,width=7.5cm}
\hspace{0.25cm}
\caption[]{ The order parameter $\bar \sigma/\Lambda$ as a function of
  $\mu /\Lambda$ for $eB=0.5\, \Lambda^2$ at $T=0$. The figure
  illustrates the discontinuities observed in Fig. 7 for the case
  $\lambda <0$.  Within the OPT the range of  $\mu$ values for which
  the intermediate global minimum happens is about one-third of the
  interval  predicted by the LN approximation.}
\label{multisigma}
\end{figure}

At low magnetic fields ($eB \lesssim \Lambda^2$) and when $\lambda
<0$,  a structure of intermediate phase transitions, where the vacuum
expectation value of the chiral condensate can jump discontinuously
from a value $\bar{\sigma}_1 \neq 0$ to another value $\bar{\sigma}_2
\neq 0$, with $\bar{\sigma}_1 > \bar{\sigma}_2$ is possible, as shown
in {}Fig. \ref{multiplePT}, where we show the normalized effective
potential for both the LN and OPT cases, ${\bar V}_N$, where

\begin{equation}
{\bar V}_N = \frac{V_{\rm eff}(\sigma_c,T=0) - V_{\rm
    eff}(\sigma_c=0,T=0)}{ N \Lambda^3}\;.
\end{equation}
Note that the multiple transitions can happen in the LN case and also
in the OPT case. Therefore, this is not an artifact of the MFA.  The
final transition is the actual chiral phase transition, where the
system jumps from $\bar{\sigma}_2 \ne 0$ to $\bar{\sigma}=0$.  In the
LN case, there is an intermediate (nonchiral) transition at a value
of critical chemical potential given by $\mu \simeq 0.878 \Lambda$
when $eB=0.5 \Lambda^2$, while in the OPT case (for $N=2$), this first
transition happens at $\mu\simeq 1.02 \Lambda$.  The actual chiral
phase transition happens at a larger value of chemical potential,
given by $\mu \simeq 1.063 \Lambda$ in the LN case and by $\mu \simeq
1.09 \Lambda$ in the OPT case.   In {}Fig.~\ref{multisigma} we show
how the chiral order parameter changes with the chemical potential,
also evidencing the intermediate transitions.
 
An analogous structure of multiple phase transitions was first
identified  in Ref.~\cite{2Bfield}, for which besides including the
perpendicular magnetic field  component, it was also considered the
inclusion of  a parallel component for the magnetic field, which
produces an enhancement of the Zeeman  energy term and an effective
spin polarization of the system. We see here that, even in the absence
of a parallel component of the magnetic field, we can also find a
similar structure. It is quite surprising that no such structure has
been reported before in the earlier literature of the GN model in a
magnetic field. {}Finally, it is interesting to note that within the
OPT the range of  $\mu$ values for which the global minimum happens at
${\bar \sigma}_2$ is about one-third of the interval predicted by the
LN approximation.

\begin{figure}[htb]
  \vspace{0.5cm}  \epsfig{figure=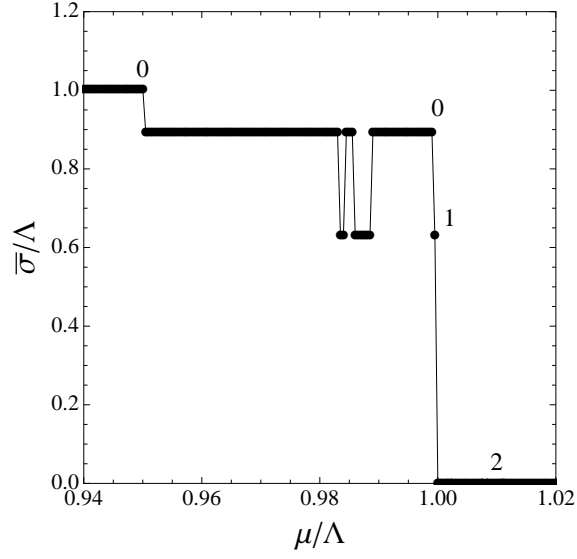,angle=0,width=7.5cm}
\hspace{0.25cm}
\caption[]{ The order parameter $\bar \sigma/\Lambda$ as a function of
  $\mu /\Lambda$ for $eB=0.2\, \Lambda^2$ at $T=0$ at large $N$ for
  $\lambda <0$. The figure illustrates the many discontinuities
  associated with the de Haas--van Alphen oscillations. The numbers
  indicate which is the highest filled Landau level for each $\bar
  \sigma$. After the transition to $\bar \sigma=0$, the levels higher
  than $j=2$ can be filled by increasing  $\mu$.}
\label{dHvA}
\end{figure}

In order to better understand these transitions we offer
{}Fig.~\ref{dHvA},  which shows the many discontinuities associated
with the de Haas--van Alphen oscillations,  which are produced when
more Landau levels are filled as $\mu$ increases. The numbers
represent the highest Landau level  which for large $N$, at $T=0$, is
given by

\begin{equation}
J_{max} = \frac{ \mu^2 - {\bar \sigma}^2}{2eB} \,\,\,,
\end{equation}
or the nearest integer.  The first transition occurs from $J_{max}=0$
to $J_{max}=0$ and is produced when  $\mu > {\bar \sigma}$, turning on
all the Heaviside step functions that appear in the free energy  at
$T=0$. The subsequent transitions occur when a different $J_{max}$ is
reached.  {}For $eB \ge 0$ the only transition is from $J_{max}=0$ to
$J_{max}=0$, since this magnetic field (which happened to be of the
order of the gap value) is high enough so that only the LLL is always
occupied. {}For $ \, \Lambda^2 > eB \gtrsim 0.4 \Lambda^2 $ there is a
transition from $J_{max}=0$ to $J_{max}=0$ associated with $\theta
(\mu - {\bar \sigma}) =1$ and then a second associated with $J_{max}=0
\to 1$. Exactly this type of behavior is observed in
    {}Fig.~\ref{multisigma} for the case of $eB=0.5 \, \Lambda^2$,
    which can now be better understood. Then, for $0.4 \, \Lambda^2 >
    eB \ge 0.2 \Lambda^2 $ we observe the type of behavior shown in
    {}Fig.~\ref{dHvA} with the transition that restores chiral
    symmetry happening at $J_{max} =2$. {}For $ 0 < eB \le 0.2
    \Lambda^2$ there are even more transitions since more levels can
    be filled. {}For example, at $eB=0.1 \Lambda^2$ chiral symmetry is
    completely restored ($\bar \sigma=0$) when $J_{max}=5$. Thus, in
    summary, for all $eB$ values a first transition occurs due to a
    nonvanishing value of $\theta (\mu - {\bar \sigma})$. If $eB \ge
    \Lambda^2$ this is the only transition and only the LLL is always
    filled, producing a smooth behavior for $\bar \sigma$ and
    $\mu_c$. In the range $0< eB < \Lambda^2$ there is also a first
    discontinuity in the value of $\bar \sigma$ due to a nonvanishing
    value of $\theta (\mu - {\bar \sigma})$, but then, since $eB$ is
    small, there can be subsequent discontinuities in $\bar \sigma$
    due to the jumps among the integer values of $J_{max}$, which
    accounts for the oscillations and discontinuities we have observed
    at $\lambda <0$ for small $eB$. When $\lambda >0$ and $eB$, the
    small chiral symmetry is restored only due to $\theta (\mu - {\bar
      \sigma})=1$ and the filling of higher Landau levels only occurs
    after $\bar \sigma=0$ by increasing $\mu$.
    Appendix~\ref{oscillations} shows how these oscillations can be
    further understood by means of Poisson's summation formula.

%%%%%%%%%%%%%%%%%%%%%%%%%%%%%%%%%%%%%%%%%%%%%%%%%%%%%%%%%%%%%%%%%%%%%%%%

\subsection{$T \ne 0$ and $ \mu \ne 0$ case}

{}Finally, in the case of finite temperature and chemical potential,
we have the effective potential as given by Eq.~(\ref{VeffHTmu}). In
this case, we search for points in the phase diagram in the plane
$(T,\mu)$, corresponding to either a first-order or a second-order
phase transition.  Recall that in a first-order transition the
effective potential develops different minima, $\bar{\sigma}^{(1)}
\neq \bar{\sigma}^{(2)}$, where one of them is a local minimum
associated with metastability, while the other is a global
minimum. These minima can get degenerate for some values of the
parameters.  {}For a given value of the magnetic field the first-order
transition points in the $(T,\mu)$ plane can be determined from the
condition of degeneracy of the minima of the effective potential, 

\begin{equation}
V_{\rm eff}(\bar{\sigma}^{(1)},B,T_c,\mu_c) = V_{\rm
  eff}(\bar{\sigma}^{(2)},B,T_c,\mu_c)\;.
\label{1stline}
\end{equation}
One of the minima is, in general, the trivial solution,
$\bar{\sigma}=0$, which then facilitates the determination of the
first-order transition points. However, as noticed in the previous
subsection, at low magnetic fields $eB \lesssim \Lambda^2$, other
minima can emerge and, thus, at low magnetic fields the determination
of the transition points must be done with care.
  
In a first-order phase transition we then have  that the minima
$\bar{\sigma}$ change discontinuously  at the transition point.  On
the other hand, the second-order phase transition critical points are
found when the nontrivial minimum $\bar{\sigma}$ changes continuously
and vanishes at the transition point.  The point where the
second-order transition line meets the first-order one, defines a
tricritical point.  The second-order and tricritical points are mostly
easily found by  using a Landau expansion for the effective potential,
which is valid for small values of the order parameter. This is the
case close to a second-order or tricritical point.  The Landau's
expansion (for small $\sigma_c$) for $V_{\rm eff}$ can be expressed in
the general form

\begin{equation}
V_{\rm eff}(\sigma_c,B,\mu,T) \simeq V_0 + \frac{1}{2} a(B,\mu,T)\:
\sigma_c^2 + \frac{1}{4} b(B,\mu,T) \:\sigma_c^4 + \frac{1}{6}
c(B,\mu,T)\: \sigma_c^6 \;,
\label{landau}
\end{equation}
where $V_0$ is a constant (independent of the order parameter)  energy
term.  Note that only even powers of $\sigma_c$ are allowed due to the
original chiral symmetry of the model.  The coefficients $a,b$, and $c$
appearing in Eq.~(\ref{landau}) can be obtained, respectively, by a
second, fourth, and sixth derivative of the effective potential
expansion around $\sigma_c=0$. Higher-order terms in the expansion
(\ref{landau}) can be verified to be much smaller than the first-order
terms and can then be consistently neglected.  In particular, note
that a tricritical point can emerge whenever we have three phases
coexisting simultaneously. 

{}From Eq. (\ref{landau}), a second-order phase transition follows
when the coefficient of the quadratic term vanishes  ($a=0$) and $b>0,
c >0$.  A first-order transition happens for the case of  $b < 0, c >
0$. The tricritical point is found when both the quadratic and quartic
coefficients in Eq.~(\ref{landau}) vanish, $a=b=0$ (with $c>0$).
Thus, Eq.~(\ref{landau}) offers a simple and immediate way for
analyzing the phase structure of our model.  {}For instance, to obtain
$T_c$ at $\mu=0$ one only needs to consider Eq.~(\ref {landau}) to
order $\sigma_c^4$ with $b>0$ to assure that the potential is bounded
from below. Then, the solution of $a(B,0,T_c)=0$ sets the critical
temperature.  However, in order to use Landau's expansion we must have
$V_{\rm eff}$ in terms of $\sigma_c$, $\mu$ and $T$ only (apart from
$N$ and the scale $\Lambda$, of course). In principle, this can be
done by using the PMS relation, Eq.~(\ref {pmsselfconsistent}).   Even
though at finite $N$, $\bar \eta$ depends on $\sigma_c$ in a highly
nonlinear way, Eq.~(\ref{pmsselfconsistent}) can be easily solved
numerically by iteration in a very efficient way (see, e.g.,
Ref.~\cite{plbgn3d}).  {}For example, at the first iteration, the use
of the approximate PMS solution obtained by using the large-$N$
solution  $\eta=\sigma_c$ within higher-order ${\cal O}(\lambda/N)$
terms,

\begin{equation}
\bar{\eta} \simeq \sigma_c + \frac{\lambda}{2 N}  \left[
  \frac{\partial I_1}{\partial \eta} + \frac{ \left(\frac{\partial
      I_1}{\partial \mu}\right) \left(\frac{\partial^2 I_1}{\partial
      \eta \partial \mu} \right)} { \frac{\partial^2 I_1}{\partial
      \eta^2} } \right]\Bigr|_{\eta = \sigma_c}\;,
\label{pmsiteration}
\end{equation}
is already able to produce results for the tricritical points within a
less than $1\%$ difference with respect to a full numerical
calculation. {}Furthermore, if the second term inside the  square
brackets in Eq.~(\ref{pmsiteration}) involving the variation of the $I_1$
term with respect to the chemical  potential is much smaller than the
first term, it can be neglected  and this can make the PMS calculation
procedure much simpler.  In all cases we have checked the
applicability of the use of this simplified form compared with the
complete expression (\ref{pmsiteration}) and used it whenever
possible to simplify the numerical calculations.  {}Following this
procedure, we obtain the tricritical  point as a function of the
magnetic field. 

\begin{figure}[htb]
  \vspace{0.5cm} 
\epsfig{figure=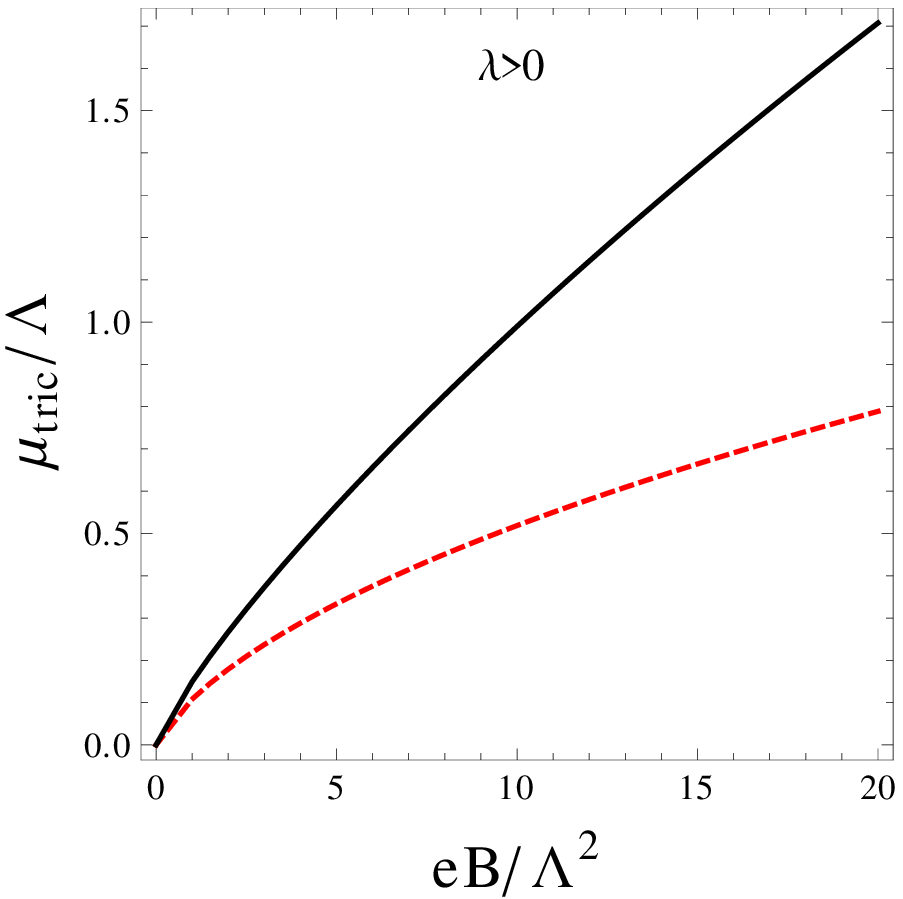,angle=0,width=7cm}
\epsfig{figure=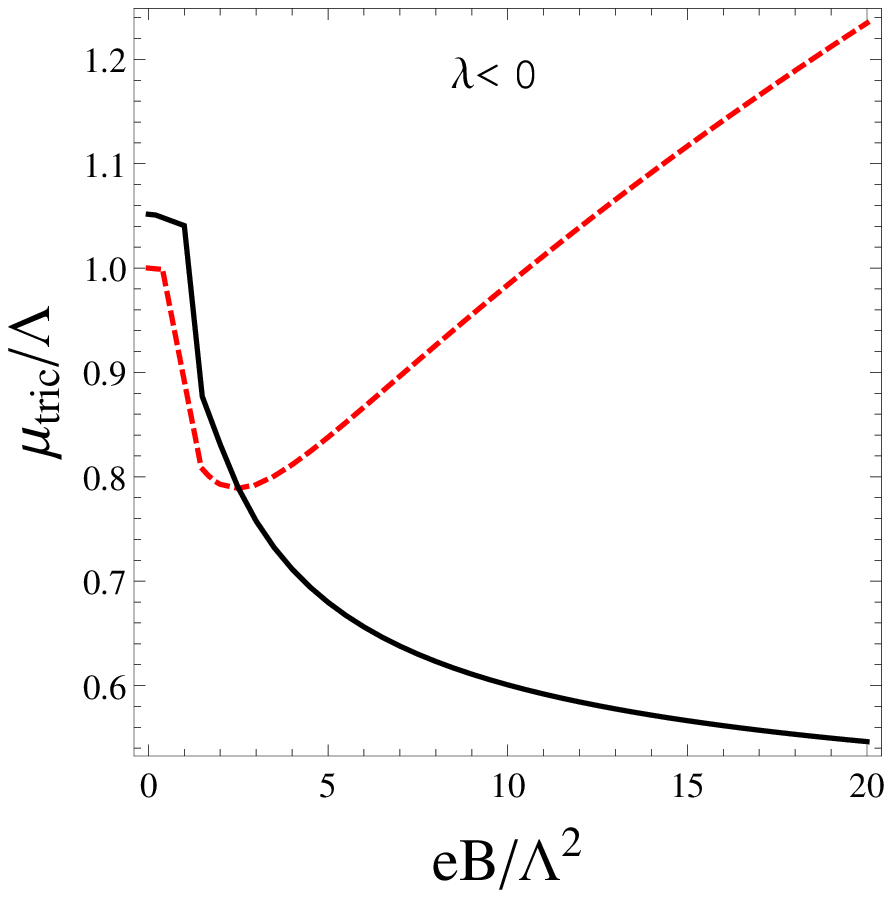,angle=0,width=7cm}
\caption[]{The chemical potential at the tricritical point for
  $\lambda >0$ (the plot on the left) and for $\lambda<0$ (the plot on
  the right), for the  LN (dashed line) and for the OPT (solid line)
  with $N=2$. }
\label{mutricfig}
\end{figure}

In {}Fig.~\ref{mutricfig} we give the results for the chemical
potential as  a function of the magnetic field at the tricritical
point, for the cases of negative and positive couplings, for both the
LN and OPT (at $N=2$) cases.  Interestingly enough, the LN results for
$\lambda <0$ (right panel) display exactly  the same qualitative
behavior found in Ref.~\cite{prc}, where the MFA was applied to the
three-flavor NJL model in 3+1 dimensions.  Note again from this figure
the effect of the IMC, similar to the one seen in {}Fig.~\ref{figmuc}.
The figure shows that $\mu_{\rm tric}$ in the OPT case only decreases
with $B$.  Again, as in the $T=0$ case  discussed in
Sec.~\ref{T0section}, this result could be a sign of the importance
brought in by the OPT $\lambda \langle \psi^\dagger \psi\rangle^2/N$
type of corrections,  which start to play an important role in this
region of intermediate to large charge asymmetries.   {}For $\lambda
>0$ the right panel of {}Fig.~\ref{mutricfig}  shows that both the OPT
and the LN approximation predict that  $\mu_{tric}$  always increase
with $B$ and that the OPT predicted values are always higher than the
LN ones.

\begin{figure}[htb]
  \vspace{0.5cm} 
\epsfig{figure=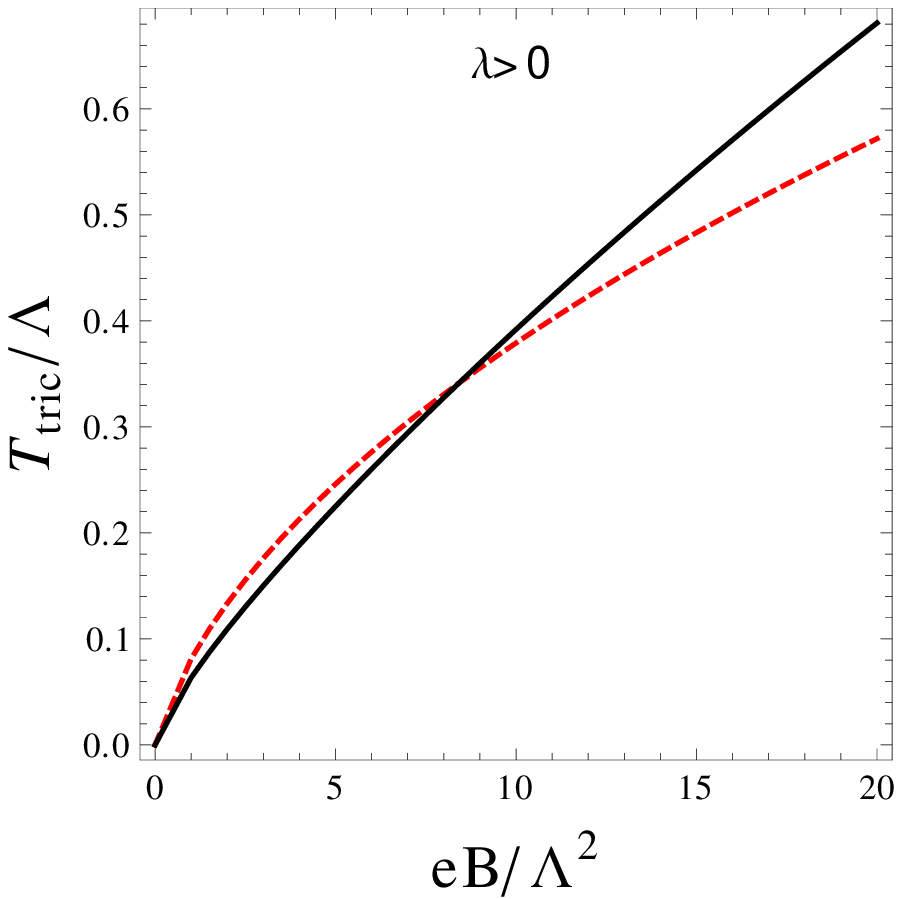,angle=0,width=7cm}
\epsfig{figure=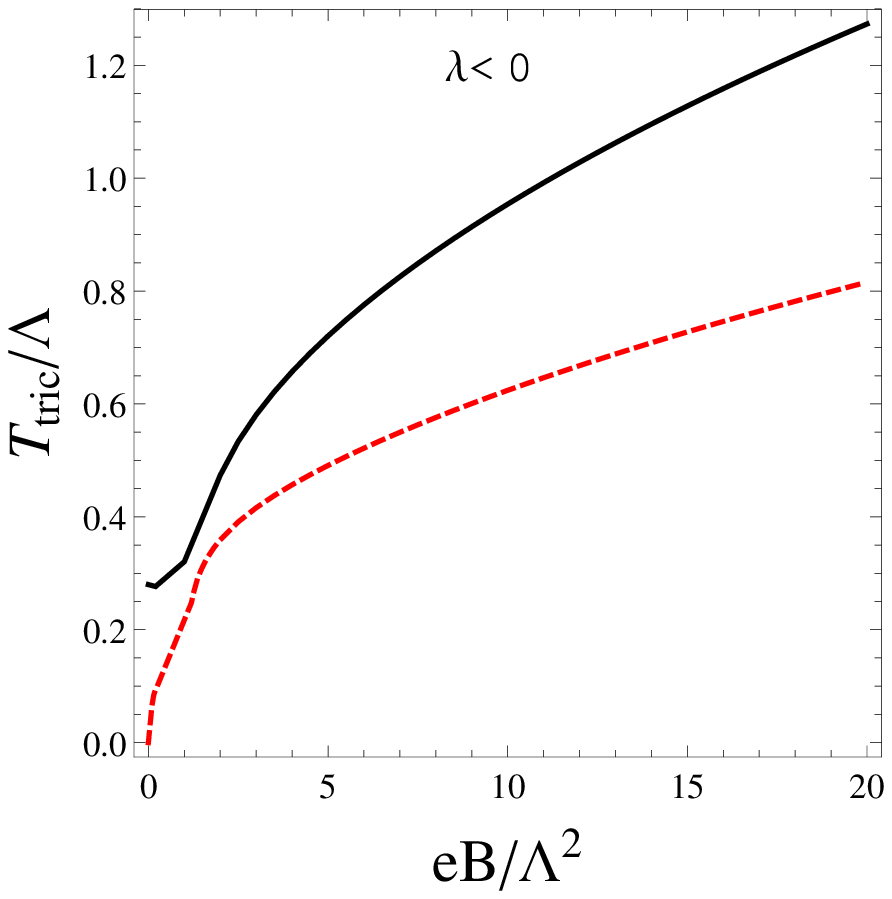,angle=0,width=7cm}
\caption[]{The temperature at the tricritical point for $\lambda >0$
  (the plot on the left) and for $\lambda<0$ (the plot on the right),
  for the  LN (dashed line) and for the OPT (solid line) with $N=2$. }
\label{Ttricfig}
\end{figure}

In {}Fig. \ref{Ttricfig} we give the results for the temperature at
the tricritical point as a function of the magnetic field,  for the
cases of negative (right panel) and positive couplings (left panel),
for both the LN and OPT (at $N=2$) cases. Let us start by discussing
the case of negative coupling at vanishing magnetic field, where the
figure shows that the LN predicts $T_{\rm tric}=0$, while the OPT
predicts $T_{\rm tric} \simeq 0.28 \, \Lambda$. As we have already
discussed (see also  Refs.~\cite {prdgn3d,plbgn3d}) this LN result for
$B=0$ can be shown to be  wrong due to  universality arguments,  while
the OPT predicted values for finite $N$ are within the  range
estimated (but not pinpointed) by Monte--Carlo
simulations~\cite{kogut}.  Despite these important quantitative
differences, both approximations show that  $T_{\rm tric}$  increases
with the magnetic field.

\begin{figure}[htb]
  \vspace{0.5cm} 
\epsfig{figure=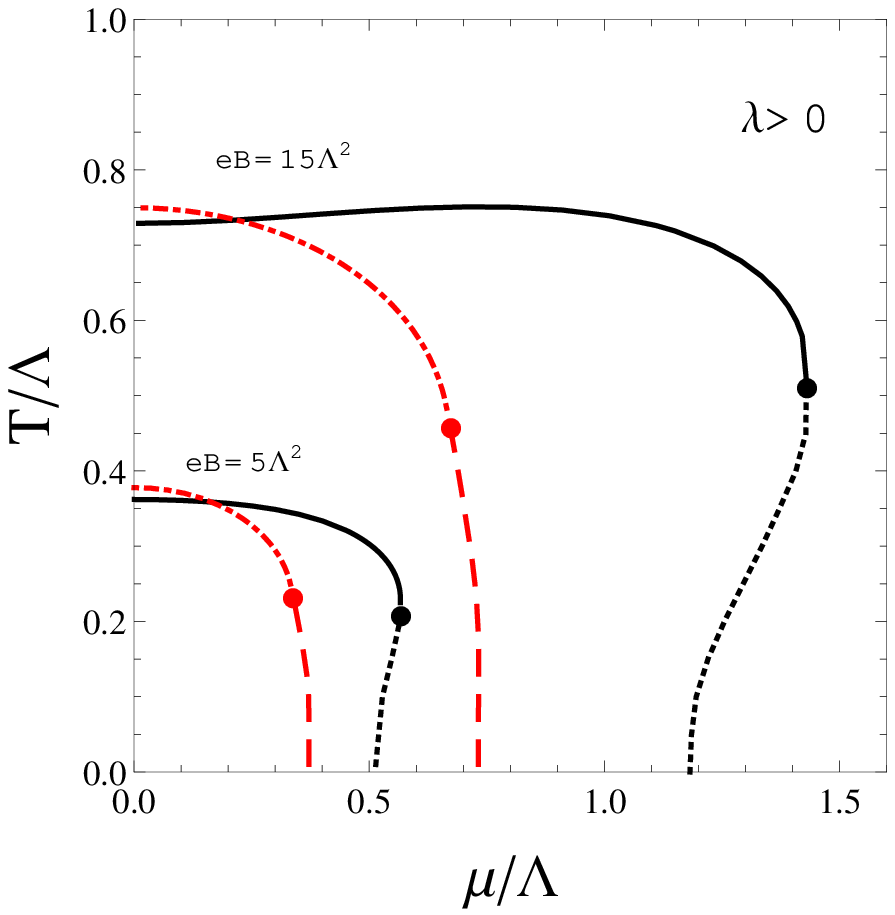,angle=0,width=7cm}
\epsfig{figure=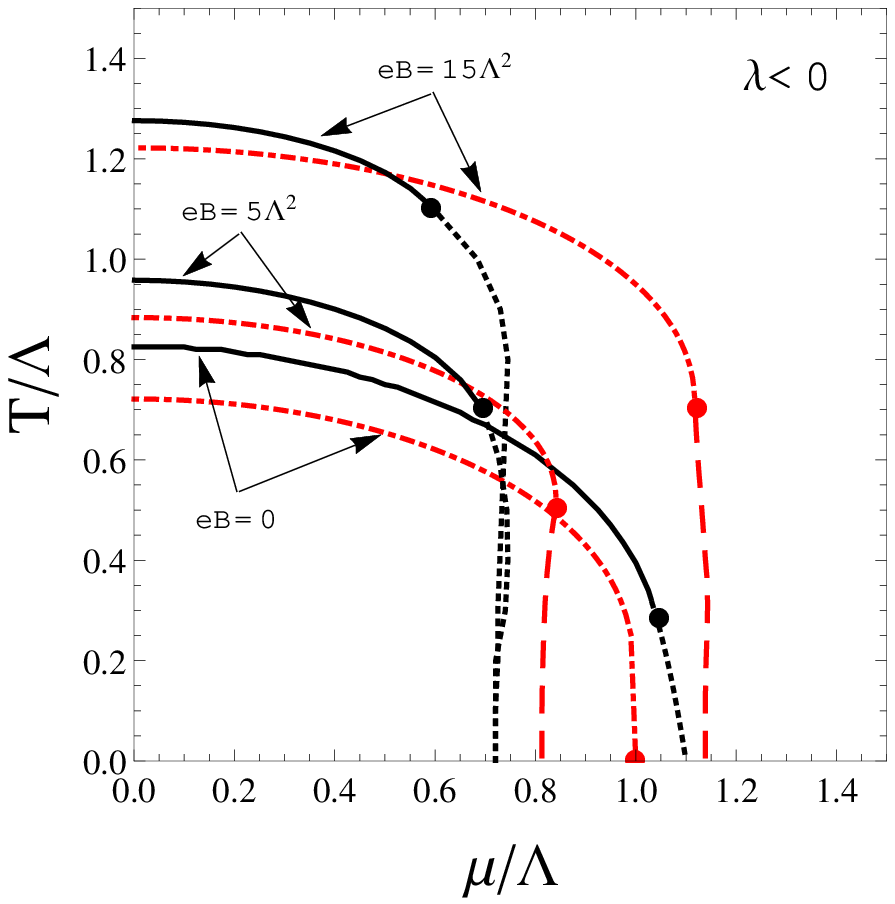,angle=0,width=7cm}
\caption[]{The phase diagram for $\lambda >0$ (the plot on the left)
  and for $\lambda<0$ (the plot on the right).  All cases are for
  $N=2$.  The solid lines indicate second-order phase transition lines
  in the OPT case, while dashed-dotted lines in the LN case. The
  first-order lines are indicated by dotted lines in the OPT case and
  by long dashed lines in the LN case.  Where the second-order line
  meets the first-order line (the tricritical point) is indicated by
  the large dot.}
\label{phasediagram}
\end{figure}

Next, in {}Fig.~\ref{phasediagram}, we show the complete phase diagram
for the LN and OPT cases for representative values of the magnetic
field. Note that in the LN case for $B=0$ and negative coupling, there
is only a second-order transition line in the $(\mu,T)$ phase diagram,
with the exception of the pair $T=0$ and  $\mu= \Lambda$, which
correspond to a first-order transition point. As shown in
Ref.~\cite{prdgn3d}, it is only by including beyond mean-field effects
that a first-order transition line (along with the tricritical point)
emerges, in agreement with the expectations based on  the results for
the GN model in 1+1 dimensions and also for the NJL model in 3+1
dimensions.  Note, however, that for nonvanishing magnetic fields, a
tricritical point is produced even in the LN case. {}For positive
couplings, recall that a chiral phase transition is only possible for
nonvanishing magnetic fields~\cite{Klimenko}.

{}Figure \ref{phasediagram} shows that for both a negative or a
positive coupling the  presence of a magnetic field  always increases
the size of the  first-order transitions. In the OPT case this
increase is even more pronounced since the term $\lambda\langle \psi^+
\psi \rangle^2/N$ enhances this type of transitions when $\lambda <0$
and the $1/N$ correction acts as an attractive vector term
\cite{fukushima}.  In both the OPT and LN cases and for any sign for
the coupling, the CSB region tends to get larger as $B$ grows.

Note that there is magnetic catalysis in $T_c$ and inverse catalysis
in $\mu_c$ for $\lambda < 0$, but  only catalysis for $\lambda > 0$ in
the OPT case. {}For the LN case and $\lambda <0$, the inverse magnetic
catalysis only happens untill some value of $B$ and then there is only
catalysis beyond that value of magnetic field. So for $eB \gtrsim 5
\Lambda^2$ or so, the CSB region in the LN case will become always
larger than in the OPT when $\lambda < 0$, while it is the opposite
when $\lambda >0$,  where $\lambda \langle \psi^+ \psi \rangle^2/N$
acts as a repulsive vector term,  which competes with the effect of
the magnetic field by enhancing more the CSB region in the OPT case
than in the LN case.

%%%%%%%%%%%%%%%%%%%%%%%%%%%%%%%%%%%%%%%%%%%%%%%%%%%%%%%%%%%%%%%%%%%%%%%

\section{Conclusions}
\label{sec5}

In this work we have analyzed the effects of a nonvanishing constant
magnetic field (applied perpendicularly  to the plane of the system)
on the phase structure of the massless discrete (2+1)-dimensional GN
model,  including contributions which  go beyond the LN (or mean
field) approximation.  Here we have used the OPT method, which has
already been successfully used before to study the properties of this
model in the absence of a magnetic field.  Both the cases of
positive and negative four-fermion coupling have been studied.

We have produced some novel results concerning the phase structure of
the model in the presence of a magnetic field.  {}For negative
couplings, we have  shown that at low magnetic fields ($eB \lesssim
\Lambda^2$) a rich structure of phases can emerge. In this case, it is
possible to have intermediate transitions to nonvanishing values of
the chiral vacuum expectation value. These transitions happen in the
LN case, something that has not been previously noted in the
literature\footnote{Recently, in Ref.~\cite{2Bfield} a similar
  structure of phase transitions has also been found, though in that
  reference it was also included  a parallel component for the
  magnetic field, which is then able to further enhance these
  intermediate phases and also to produce reentrant phase
  transitions. However, in the absence of the perpendicular component
of the magnetic field, the intermediate and reentrant phases are
both absent~\cite{zeeman}.}  and remains also when including corrections beyond
the LN approximation, as we have shown by using the OPT
method. Therefore, these intermediate transitions are not an artifact
of the LN approximation. We have also traced the origin of these
intermediate transitions as being a consequence of the magnetic de
Haas--van Alphen oscillations that arise at low values of the magnetic
field, $e B \lesssim \Lambda^2$ and for negative values of the
coupling constant.

As for the effect of the magnetic field on the phase structure of the
model, we have shown two distinct effects depending on the sign of the
coupling constant.  {}For either positive or negative couplings, we
still have an enhancement of the chiral-broken-symmetry region, as
expected in general from the magnetic catalysis effect.  However, when
the coupling is negative, beyond some value of the magnetic field the
chiral broken region is always smaller in the OPT case than in the LN,
while for positive coupling the reverse is observed, with the OPT
always producing a larger broken symmetry region. The tricritical
points tend to be also enhanced in general by the presence of the
magnetic field, with the results obtained in the OPT larger than in
the LN case. The exception is the value of the chemical potential at
the tricritical point, $\mu_{\rm tricrit}$ when the coupling is
negative. When $\lambda <0$,  in the LN case $\mu_{\rm tricrit}$ tends
to be strongly suppressed by the magnetic field initially, until for
$eB \gtrsim 2 \Lambda^2$ it turns again to be enhanced by the magnetic
field. This decrease of the critical chemical potential with the
magnetic field, the inverse magnetic catalysis effect, has some similarity  
with the phenomenon
seen in the  NJL model and discussed in details recently in
Ref.~\cite{andreas}. But in our case it originates from the OPT 
$\lambda\langle \psi^\dagger \psi\rangle/N$ corrections beyond LN, 
as explained in Sec.~\ref{T0section}. 
In the OPT case $\mu_{\rm tricrit}$ continues to
decrease for very  large magnetic fields. Thus, the
inverse magnetic catalysis remains unsuppressed even for  large values
of the magnetic field in the OPT context, which is opposite to what is
seen  in the LN case.  Note also that inverse magnetic catalysis is
seen to operate only on the critical chemical potential values, while
the critical temperature still shows only the standard magnetic
catalysis, always increasing with $B$. This same trend also applies to
the coexistence chemical potential when $T=0$, where, for $\lambda < 0$, 
the OPT shows an inverse magnetic catalysis effect even for large magnetic 
fields, while in the LN case, the chemical potential only decreases for 
relatively small values of the magnetic field and grows for 
$eB \gtrsim 2 \Lambda^2$. 
At $\mu=0$, the critical temperature is seen only to increase with the
magnetic field in both the OPT and LN cases, independently of the sign of
the coupling constant. The stability of the order-$\delta$ results for the 
same model, at $B=0$, has been addressed in Ref. \cite {prdgn3d} and the 
outcome of that investigation allows us to believe that our present results, 
at $B \ne 0$, should also be stable against the inclusion higher-order corrections.

It is tempting to compare our results with recent lattice results for
the QCD chiral crossover temperature as function of $B$. 
The first lattice studies~\cite {earlylattice} considered two quark flavors, 
with high values of pion masses ($m_\pi=200-400 \, {\rm MeV}$), and have shown 
that the critical temperature should increase with $B$. 
However, an improved lattice simulation \cite {Bali:2011qj}, which considered 
2+1 quark flavors at physical pion mass values ($m_\pi=140 \, {\rm MeV}$), 
together with an extrapolation to the continuum, predicted that  
the critical temperature should decrease with $B$. 
Since then, most models have tried to reproduce these lattice results,
showing, however, that the critical temperature only increases with the magnetic
field. 
Since most model results were obtained within the LN/MFA, one may wonder if the 
discrepancy could not be resolved by going beyond this approximation.
Our results indicate that this may not be sufficient and that other effects 
besides going beyond the LN approximation are required.  

We hope that our findings will give further insights in applications
that employ four-fermion models in the description of planar condensed
matter systems, which we intend to further explore in the future.

\appendix

\section{Summing Matsubara frequencies, Landau levels and related formulas}
\label{definitions}

Let us derive here the momentum integrals appearing in the expression
for the effective potential (\ref{vlde1ddim}) and then give the $I_i$,
$i=1,2,3$, integrals  Eqs. (\ref{int1}), (\ref{int2}) and
(\ref{int3}).  Using the replacements (\ref{intp}), $p_0 \to
i(\omega_\nu - i \mu)$ and ${\bf p}^2 \to 2 j eB$, with $\omega_\nu=
(2 \nu+1)\pi T$, $\nu=0,\pm 1,\pm 2,\ldots$, are the Matsubara
frequencies for fermions and $j$ labels the LLs. The
integral $I_1$ is defined as

\begin{eqnarray}
I_1 &=& i\int_p {\rm ln} (P^2 - \eta^2 )  \nonumber \\ &=& -T
\sum_{\nu=-\infty}^{+\infty} \frac{eB}{4\pi}\sum_{j=0}^{\infty}
\alpha_j \ln [(\omega_{\nu}-i \mu)^2 + E_j^2]
\label{intln}
\end{eqnarray}
where  $E_j=\sqrt{2jeB + \eta^2}$ and $\alpha_j= 2-\delta_{j, 0}$.
Performing the  Matsubara sum one gets

\begin{equation}
I_1= -\frac{eB}{4\pi}\sum_{j=0}^{\infty}\alpha_j\left\{ E_j + T \ln[1+
  e^{-(E_j+\mu)/T}] + T \ln [1+ e^{-(E_j-\mu)/T}] \right \}\;.
\label{intI1}
\end{equation}

In the same way, we have that

\begin{eqnarray}
I_2 &=& i\int_p \frac{1}{P^2 - \eta^2}  \nonumber \\ &=&
\frac{eB}{8\pi}\sum_{j=0}^{\infty}\alpha_j \left \{\frac {1}{ E_j} -
\frac {1}{ E_j [1+ e^{(E_j+\mu)/T}]} - \frac{1}{E_j  [1+
    e^{(E_j-\mu)/T}]} \right \}\;,
\label{intI2}
\end{eqnarray}
while the last momentum integral remaining that we need is

\begin{eqnarray}
I_3 &=& - i \int_p \frac{P_0}{P^2- \eta^2 } \nonumber \\ &=&
\frac{eB}{8\pi}\sum_{j=0}^{\infty}\alpha_j \left
     [\frac{\sinh(\mu/T)}{\cosh(\mu/T)+\cosh(E_j/T)} \right ]
     \nonumber \\ &=& \frac{eB}{8\pi}\sum_{j=0}^{\infty}\alpha_j\left[
       \frac{1}{e^{\left( E_j-\mu \right) /T}+1}-\frac{1}{e^{\left(
           E_j+\mu \right) /T}+1}\right]\,.
\label{intI3}
\end{eqnarray}

There is a very convenient trick to perform the sum over Landau levels
for the $T,\mu$ independent terms which can be expressed in a closed
form by means of Riemann--Hurwitz zeta functions~\cite{zeta}. {}For
example, consider the $T$ \and $\mu$ independent term on the 
right-hand side of Eq. (\ref{intI1}). By adding and subtracting a lowest
Landau energy level term, $E_0$, to it one can write

\begin{equation}
-\frac{eB}{4\pi}\sum_{j=0}^{\infty}\alpha_j E_j
+\frac{eB}{4\pi}E_0-\frac{eB}{4\pi}E_0=  \frac{eB |\eta|}{4\pi} -
\frac{(2eB)^{3/2}}{4\pi} \sum_{j=0}^{\infty} \left [ j +
  \frac{\eta^2}{2eB} \right ]^{1/2} \,\,.
\end{equation}
The infinite sum can be related to the Riemann--Hurwitz zeta function
Eq.~(\ref{zetafunc}), yielding

\begin{equation}
-\frac{eB}{4 \pi}\sum_{j=0}^{\infty}\alpha_j E_j= 
\frac{eB|\eta|}{4\pi}- \frac{(2eB)^{3/2}}{4\pi} 
\zeta\left(-\frac{1}{2},\frac{\eta^2}{2eB} \right )\,.
\end{equation}

The same technique, when applied to the $T$ and $\mu$ independent term
of Eq. (\ref{intI2}), gives

\begin{equation}
\frac{eB}{8\pi}\sum_{j=0}^{\infty}\alpha_j  \frac {1}{ E_j}= -
\frac{eB}{8 \pi |\eta|} + \left (2eB\right )^{1/2}\frac{1}{8\pi}
\zeta\left (\frac{1}{2},\frac{\eta^2}{2eB} \right )  \,\,.
\end{equation}
The same types of manipulations can be applied to divergent terms, as
discussed in Refs. \cite{prc,andre}.

It is also useful to have the limiting cases for the functions $I_i$
when $T=0$ and/or $\mu=0$. {}Taking the $T=0$ limit in
Eqs. (\ref{intI1}), (\ref{intI2}) and (\ref{intI3}), we obtain

\begin{eqnarray}
&& I_1 (\eta, B, T,\mu) \stackrel{T\to 0}{\longrightarrow} \frac{eB}{4
    \pi} |\eta| - \frac{(2 eB)^{3/2}}{4 \pi} \zeta\left( -
  \frac{1}{2}, \frac{\eta^2}{2 eB} \right) - \frac{eB}{4 \pi}
  \sum_{j=0}^{J_{\rm max}} \alpha_j \left( \mu - E_j \right)
  \theta(\mu-|\eta|)\;,
\label{I1T0}
\\ && I_2 (\eta, B, T,\mu) \stackrel{T\to 0}{\longrightarrow}
-\frac{eB}{8 \pi |\eta|}  + \frac{(2eB)^{1/2}}{8\pi} \zeta\left
(\frac{1}{2},\frac{\eta^2}{2eB} \right)- \frac{eB}{8 \pi}
\sum_{j=0}^{J_{\rm max}} \alpha_j \frac{1}{E_j} \theta(\mu-|\eta|)\;,
\label{I2T0}
\\ && I_3 (\eta, B, T,\mu) \stackrel{T\to 0}{\longrightarrow}
\frac{eB}{8 \pi} \theta(\mu-|\eta|) + \frac{eB}{4 \pi}  {\rm
  Int}\left( \left| \frac{\mu^2 - \eta^2}{2 eB} \right| \right)
\theta(\mu-|\eta|)\,,
\label{I3T0}
\end{eqnarray}
where in the above equations $J_{\rm max}= {\rm Int}((\mu^2 -
\eta^2)/(2 eB))$ and ${\rm Int}(x)$ means the integer part of $x$.

{}Taking the $\mu=0$ limit in  Eqs. (\ref{intI1}), (\ref{intI2}) and
(\ref{intI3}), we obtain

\begin{eqnarray}
&& I_1(\eta,B,T,\mu) \stackrel{\mu\to 0}{\longrightarrow} \frac{eB
    |\eta|}{4\pi} - \frac{(2eB)^{3/2}}{4\pi} \zeta
  \left(-\frac{1}{2},\frac{\eta^2}{2eB} \right)
  -\frac{eBT}{2\pi}\sum_{j=0}^{\infty}\alpha_j \ln\left(1+
  e^{-E_j/T}\right) \;,
\label{I1mu0}
\\ && I_2 (\eta, B, T,\mu) \stackrel{\mu\to 0}{\longrightarrow} -
\frac{eB}{8 \pi |\eta|} +  \frac{(2eB)^{1/2}}{8\pi} \zeta\left
(\frac{1}{2},\frac{\eta^2}{2eB} \right ) -
\frac{eB}{4\pi}\sum_{j=0}^{\infty}\alpha_j \frac{1}{ E_j \left(1+
  e^{E_j/T}\right) } \;,
\label{I2mu0}
\\ && I_3 (\eta, B, T,\mu) \stackrel{\mu\to 0}{\longrightarrow} 0\;,
\label{I3mu0}
\end{eqnarray}
while for $T=0$ and $\mu=0$, we obtain

\begin{eqnarray}
&& I_1(\eta,B,T,\mu) \stackrel{T,\mu\to 0}{\longrightarrow} \frac{eB
    |\eta|}{4\pi} - \frac{(2eB)^{3/2}}{4\pi} \zeta
  \left(-\frac{1}{2},\frac{\eta^2}{2eB} \right) \;,
\label{I1Tmu0}
\\ && I_2 (\eta, B, T,\mu) \stackrel{T,\mu\to 0}{\longrightarrow} -
\frac{eB}{8 \pi |\eta|} +  \frac{(2eB)^{1/2}}{8\pi} \zeta\left
(\frac{1}{2},\frac{\eta^2}{2eB} \right) \;,
\label{I2Tmu0}
\\ && I_3 (\eta, B, T,\mu) \stackrel{T,\mu\to 0}{\longrightarrow} 0\;.
\label{I3Tmu0}
\end{eqnarray}

%%%%%%%%%%%%%%%%%%%%%%%%%%%%%%%%%%%%%%%%%%%%%%%%%%%%%%%%%%%%%%%

\section{Low-$B$ behavior and de Haas--van Alphen oscillations}
\label{oscillations}

It is instructive to analyze the origin of the oscillating behavior at
low $B$ found in the case of $T=0$ and $\lambda <0$ in
Sec. \ref{T0section}.  Since the terms $I_2$ and $I_3$ are directly
derived from $I_1$, it is enough for our purposes here to analyze the
low-$B$ behavior of only $I_1$.  The $\mu \neq 0$ dependent part of
$I_1$ at $T=0$, from Eq.~(\ref{I1T0}), is given by

\begin{eqnarray}
 I_1^{\mu \neq 0, T=0} &=& -\frac{eB}{4 \pi} \sum_{j=0}^{J_{\rm max}}
 \alpha_j \left( \mu - E_j \right) \theta(\mu-|\eta|) \nonumber \\  &
 = & -\frac{eB}{4 \pi} \sum_{j=0}^{\infty} \alpha_j \left( \mu - E_j
 \right) \theta(\mu - E_j)  \nonumber \\  &=& -\frac{eB}{4 \pi}
 \sum_{j=-\infty}^{\infty} \left( \mu - E_{|j|} \right) \theta(\mu -
 E_{|j|})  \nonumber \\  & = & -\frac{eB}{2 \pi}
 \sum_{n=-\infty}^{\infty}  \int_{0}^{J_{\rm max}} dy\, \left[ \mu -
   E(y) \right] \cos(2 \pi n y)\theta(\mu-|\eta|)\;,
\label{I1osc}
\end{eqnarray}
where $E(y) = \sqrt{2eB y + \eta^2}$ and in the last line in the above
equation we have made use of the  Poisson's summation
formula~\cite{poisson},

\begin{equation}
\sum_{j=-\infty}^{\infty} f(j) = \sum_{n=-\infty}^{\infty}
\int_{-\infty}^{\infty} dy f(y) e^{-2 \pi i n y}\;.
\end{equation}
Performing the integral in Eq.~(\ref{I1osc}), we find

\begin{eqnarray}
 I_1^{\mu \neq 0, T=0} &=& -\frac{1}{4\pi} \left[\frac{\mu^{3}}{3}-
   \mu \eta^{2}+\frac{2}{3}|\eta|^{3}+\frac{eB}{\pi}|\eta|
   \sum_{n=1}^{\infty}\frac{1}{n}P_{2}\left(\frac{n\pi}{eB}\eta^{2}\right)\right]
 \theta(\mu-|\eta|) \nonumber \\ 
&+& \frac{eB}{4\pi^{2}} \mu
 \sum_{n=1}^{\infty}\frac{1}{n}
 \left\{P_{2}\left(\frac{n\pi}{eB}\mu^{2}\right)\,
 \cos\left[\frac{n\pi}{eB}(\eta^{2}- \mu^{2})\right] \right.
\nonumber \\
&+& \left. Q_{2}\left(\frac{n\pi}{eB}\mu^{2}\right)\,
 \sin\left[\frac{n\pi}{eB}(\eta^{2}-\mu^{2})\right]\right\}
 \theta(\mu-|\eta|) \, ,
\label{I1osc2}
\end{eqnarray}
where the functions $P_{2}(x)$ and $Q_{2}(x)$ are defined as

\begin{eqnarray}
&&Q_{2}(x)=\sqrt{\frac{2\pi}{x}}\left\{\left[C_{2}(x)-\frac{1}{2}\right]
\cos(x)+\left[S_{2}(x)-\frac{1}{2}\right]\sin(x)\right\}\,,
\nonumber \\ 
&&P_{2}(x)=\sqrt{\frac{2\pi}{x}}\left\{\left[C_{2}(x)-\frac{1}{2}\right]
\sin(x)-\left[S_{2}(x)-\frac{1}{2}\right]\cos(x)\right\}\,,
\end{eqnarray}
and $C_2(x)$ and $S_2(x)$ are  the Fresnel integrals~\cite{Bateman},

\begin{eqnarray}
&&C_{2}(x)\equiv \frac{1}{\sqrt{2\pi}}\int_{0}^{x}dt
  \ t^{-\frac{1}{2}}\cos t\,,  \nonumber \\  
&&S_{2}(x)\equiv
  \frac{1}{\sqrt{2\pi}}\int_{0}^{x}dt \ t^{-\frac{1}{2}}\sin t \, .
\end{eqnarray}

In Eq.~(\ref{I1osc2}) the oscillatory character of the effective
potential at low-$B$  becomes explicit, as the origin of the de
Haas--van Alphen oscillations in magnetized systems. {}For $\lambda<0$
in the symmetry broken phase, as far $\mu > \eta$ and since $\eta$ is
determined from the PMS solution  (\ref{baretaT0mu}), $\bar\eta
\approx  \bar\sigma \approx \Lambda$, we have that for $e B \lesssim
\Lambda^2$ all quantities depending on $I_1$ will exhibit an
oscillatory behavior, with the different basic functions in
Eq.~(\ref{I1osc2}) of periods $\propto (e B)^{-1}$  characteristic of de
Haas--van Alphen oscillations. This determines the low-$B$ behavior
seen in {}Figs. \ref{figmuc} and \ref{figmuclowH}, as well the
multiple minima structure exhibited by the effective potential for
values of magnetic field $e B \lesssim \Lambda^2$ and shown in the
example given by {}Fig.~\ref{multiplePT}.  When $\lambda > 0$, we know
that in the absence of a magnetic field there is no chiral symmetry
breaking. However, when $B\neq 0 $, magnetic catalysis induces  chiral
symmetry breaking and, as first shown in Ref.~\cite{Klimenko}, in the
LN case, the vacuum expectation value for $\sigma$ is 
$\bar \sigma \sim  e B/\Lambda$, but we have that
$\bar \eta \sim  \bar \sigma$ and then,  for $\mu > \eta \sim \bar
\sigma$, the magnetic oscillations become highly suppressed when
$\lambda > 0$.  This is why we do not see (or hardly can see any)
oscillations in this case.

%%%%%%%%%%%%%%%%%%%%%%%%%%%%%%%%%%%%%%%%%%%%%%%%%%%%

\acknowledgments

M.B.P. and  R.O.R. are partially supported by Conselho Nacional de
Desenvolvimento Cient\'{\i}fico e Tecnol\'ogico (CNPq).  R.O.R. is also
partially supported by a research grant from  Funda\c{c}\~ao Carlos
Chagas Filho de Amparo \`a Pesquisa do Estado do Rio de Janeiro
(FAPERJ) and M.B.P. is also partially supported by   Funda\c{c}\~ao de
Amparo \`a Pesquisa e Inova\c c\~{a}o do Estado de Santa Catarina
(FAPESC).

%%%%%%%%%%%%%%%%%%%%%%%%%%%%%%%%%%%%%%%%%%%%%%%%%%%%

\end{document}